%% file: main.tex
\documentclass[
  aps,
  prb,
  amsmath,amssymb,
  preprint,
]{revtex4-2}

\usepackage[T1]{fontenc}
\usepackage[utf8]{inputenc}
\usepackage{graphicx}
\usepackage{amsfonts}
\usepackage{mathtools}
\usepackage{nicefrac}
\usepackage{physics}
\usepackage{textgreek}
\usepackage{placeins}
\usepackage{xcolor}
\usepackage{url}
\usepackage[caption=false]{subfig}
\usepackage[hidelinks]{hyperref}
\usepackage{longtable}
\usepackage{array}

\input{commands}

\begin{document}
\title[Self-consistent HL-MGGA within PAW]{Self-consistent Hessian-level meta-generalized gradient approximation}

\author{Pooria Dabbaghi}

\author{Juan Maria García Lastra}

\author{Piotr de Silva}
\email{pdes@dtu.dk}
\affiliation{Department of Energy Conversion and Storage, Technical University of Denmark, Agnes Nielsens Vej 301, 2800 Kongens Lyngby, Denmark}

\begin{abstract}
The $\vartheta$-MGGA class of density functionals is formally reformulated as Hessian-level meta-generalized gradient approximations (HL-MGGAs). 
In contrast to standard meta-GGAs that rely on the orbital-dependent kinetic-energy density or the density Laplacian, 
HL-MGGAs utilize the full density Hessian. 
We introduce a simplified, non-empirical functional, $\vartheta$-PBE, 
and present a roadmap for its self-consistent implementation within the projector augmented-wave (PAW) method. 
By utilizing the complete set of spatial second-order density derivatives, the functional's underlying descriptor successfully distinguishes 
between distinct one-electron density limits, such as single-center atomic densities and two-center bonds, that standard iso-orbital indicators often conflate. 
Benchmarks across molecular and solid-state datasets reveal that while $\vartheta$-PBE delivers accurate chemisorption energies and molecular properties, 
challenges remain in predicting bulk lattice constants. 
Ultimately, this work demonstrates the physical utility and feasibility of designing orbital-independent, Hessian-based exchange-correlation functionals.
\end{abstract}

\maketitle
\input{intro}
\input{theory}
\input{scfimp}
\input{results}
\input{summary}

\bibliographystyle{unsrt}
\bibliography{references}
\onecolumngrid
\include{appendix}

\twocolumngrid
\end{document}

%% file: commands.tex
\DeclarePairedDelimiter{\altket}{\vert}{\rangle}
\DeclarePairedDelimiter{\altbra}{\langle}{\vert}
\newcommand{\altbraket}[2]{\ensuremath{\braket{\smash{#1}}{\smash{#2}}}}

\newcommand{\functional}[2]{#1{\left[#2\right]}}
\newcommand{\function}[2]{#1{\left(#2\right)}}
\newcommand{\funcr}[1]{\function{#1}{\vrr}}
\newcommand{\vrr}{\mathbf{r}}
\newcommand{\rd}{\mathrm{d}}
\newcommand{\hessian}{\nabla^{(2)}}
\newcommand{\funcd}[2]{\frac{\partial{#1}}{\partial{#2}}}

\newcommand{\functionald}[2]{\frac{\delta{#1}}{\delta{#2}}}

\newcommand{\diff}{\mathrm{d}}
\newcommand{\funcdif}[2]{\frac{\diff{#1}}{\diff{#2}}}

\newcommand{\nrho}{{\nabla\!{n}}}
\newcommand{\nrhot}{{\nabla}^T\!{\!n}\,}

%% file: intro.tex
\section{Introduction}
Density functional theory (DFT) \cite{hk1964,ks1965} provides an exact formalism for the ground-state energy and density of interacting electrons, but in practice its accuracy depends on approximating the unknown exchange-correlation (xc) functional, $E_{\rm xc}[n]$. 
Semilocal approximations retain computational efficiency by expressing $E_{xc}$ as a single real-space integral over an xc energy density. 
Generalized-gradient approximations (GGAs) \cite{pw86, becke88, pbe1996, rpbe1999, pbesol, pbemol, pbeint} improve upon the 
local spin-density approximation (LSDA) \cite{ks1965,lsda_vonbarth1972, lda_pw1992} by enforcing additional exact constraints
according to the density gradient \cite{ge2,mabruckner}.
However, it is known that no single GGA can simultaneously satisfy all exact constraints or excel for all types of systems; 
improving accuracy in one regime often incurs deterioration in another \cite{pbesol,pbemol,pbeparams}.

Meta‐GGAs (MGGAs)
\cite{perdew85, becke1989_mgga, mgga_ghoshparr86, vs1998, tpss2003, scan2015, rtpss2018, r2scan2020, LAK2024, scanl, r2scanl, ofr2}
address this limitation by introducing one or more higher-order ingredients,
most commonly the orbital-dependent kinetic-energy density, $\tau = \frac{1}{2}\sum_i^{\text{occ}}|\nabla{\psi_i}|^2$,
the density Laplacian $\nabla^2 n$ or both.
These additional ingredients provide the flexibility required to satisfy
a broader range of exact constraints while preserving computational efficiency as the semilocal form of the functional is retained.
The meta-GGA energy density expressions often employ dimensionless region indicators that
identify local electronic characters and enable smooth interpolation between appropriate limiting behaviors.
Typical examples include the iso-orbital indicator $z = \tau^W / \tau$
and the inhomogeneity parameter $\alpha = (\tau - \tau^W) / \tau^{\mathrm{unif}}$,
where $\tau^W$ and $\tau^{\mathrm{unif}}$ denote the Weizsäcker~\cite{von1935} and Thomas-Fermi~\cite{fermi1927} kinetic energy densities, respectively.
The advantage of these region indicators has been demonstrated primarily in widely used functionals
such as TPSS~\cite{tpss2003} (using $z$) and SCAN~\cite{scan2015} (using $\alpha$), with the latter achieving remarkable accuracy for both molecular energetics and bulk properties.
Nevertheless, shortcomings remain; for example, SCAN tends to overbind adsorbates on metal surfaces,
whereas TPSS and the more recent RTPSS~\cite{rtpss2018} yield improved chemisorption energetics but
are comparatively less accurate for bulk properties.
Meta-GGAs have also shown promise in addressing the band-gap underestimation problem in DFT~\cite{LAK2024}.
These developments illustrate how the meta-GGA framework can be systematically extended 
to achieve balanced accuracy across diverse bonding regimes and material properties, 
while maintaining a non-empirical foundation.

%
Beyond standard meta-GGAs, the principle of local adaptation to the electronic character also extends
to certain GGAs~\cite{pbeint} and deorbitalized Laplacian-level meta-GGAs (LL-MGGAs)~\cite{scanl,r2scanl,ofr2}. 
In the latter class, the kinetic energy density is replaced by its explicit density-based approximations. 
These approaches demonstrate that region indicators can be constructed entirely from the electron density and its derivatives, 
bypassing the need for orbital-dependent quantities.
This distinction is of significant formal and practical importance. By retaining explicit density dependence, these functionals operate strictly within the Kohn-Sham framework, yielding a local xc potential, $v_{xc}(\mathbf{r})$. In contrast, orbital-dependent ingredients like $\tau$ typically necessitate the Generalized Kohn-Sham (GKS) formalism~\cite{gks}, which introduces computationally more demanding non-local potentials or differential operators. Consequently, explicit density functionals preserve the computational efficiency and implementation simplicity of GGAs while capturing meta-GGA-level physics.
For instance, the single exponential decay detector (SEDD)~\cite{desilva2012revealing,desilva2013extracting,sedd} was designed to identify regions where the density decays exponentially, heuristically assuming that such regions are dominated by a single orbital. 
Similarly, the density overlap region indicator ($\text{DORI} = \vartheta / (1 + \vartheta)$)~\cite{dori} 
is another $\tau$-independent quantity built from the inhomogeneity function $\vartheta$, defined as
\begin{equation}
	\label{eq:theta}
	\vartheta=\frac{\left(\nabla\left(\frac{\nabla n}{n}\right)^2\right)^2}
    {\left(\left(\frac{\nabla n}{n}\right)^2\right)^3},
\end{equation}
which has been employed in the $\vartheta$-MGGA functional~\cite{thetamgga} as the primary region indicator.
Applying the gradient operator in Eq.~(\ref{eq:theta}) yields
\begin{equation}
\label{eq:theta_expand}
\vartheta=4\!\left(1-\frac{n}{\gamma^2}\,\nabla^{T}\!n\,\nabla\gamma+\frac{n^{2}}{4\gamma^{3}}\,\nabla^{T}\!\gamma\,\nabla\gamma\right),
\end{equation}
where $\gamma=\nabla^{T}\!n\,\nabla n=\lvert\nabla n\rvert^2$ is the squared density gradient, and
$\nabla\gamma=2\,\nabla^{(2)}\!n\,\nabla n$ depends on the density Hessian $\nabla^{(2)}n$.
Since $\vartheta$ explicitly involves $\nabla^{(2)} n$, functionals that employ it are naturally classified as Hessian-level meta-GGAs (HL-MGGAs). 
Using $\nabla^{(2)}n$ offers additional flexibility for detecting different electronic regions based on the local curvature of the density 
and the variation of its gradients for constructing xc functionals. 
However, to maintain universality, the Hessian must enter in a reduced form---as in Eq.~\eqref{eq:theta_expand} through the gradient-projected Hessian---although 
this is not the only possible choice. 
Another candidate is the reduced density Hessian \cite{pig}, defined as
\begin{equation}
\label{eq:rdh}
p = \frac{\nabla^{T} n \nabla^{(2)} n \nabla n}
{4(3\pi^{2})^{2/3} |\nabla n|^{2} n^{5/3}},
\end{equation}
where Eq.~\eqref{eq:theta_expand} can be well approximated as
\begin{equation}
\label{eq:theta_approx}
\vartheta \approx 4\left(1 - p / s^{2}\right)^2,
\end{equation}
with $s = {|\nabla n|}/{2 (3\pi^2)^{1/3} n^{4/3}} $ 
denoting the reduced density gradient.
Both $p$ and $\vartheta$ have been employed as local indicators
for the calibration and construction of local hybrid functionals \cite{pig, thetahyb2015}.
Furthermore, $\vartheta$ has been tested in deorbitalization strategies with promising initial results \cite{scanl},
providing another route towards developing HL-MGGA functionals.
It is also worth noting that the (reduced) density Hessian naturally
arises in the construction of exchange-correlation potentials
at the GGA level through terms involving $\nabla s$,
as well as in the calculation of LL-MGGA xc contributions to the stress tensor, 
and is therefore not an unfamiliar quantity in DFT.


In this work, we present an implementation of the Hessian-level meta-GGA functional framework for materials simulations. As a prototype of such a functional, we introduce and validate $\vartheta$-PBE ---a modified version of the previously proposed $\vartheta$-MGGA functional. The latter was implemented in non-self-consistent post-SCF calculations and validated for molecular systems only. Therefore, the main aim of this work is to extend the formalism to self-consistent periodic calculations using the projector-augmented-wave (PAW) method~\cite{blochl-paw}. As a widely adopted generalization of pseudopotential techniques, the PAW method maps the highly oscillatory all-electron wave functions into a smooth, 
computationally tractable pseudo-space representation. 
This yields well-behaved electron densities that provide the numerical robustness required to evaluate the higher-order derivatives characteristic of HL-MGGA exchange-correlation potentials. 
Such evaluations have historically proven challenging~\cite{neumann1997higher} for the finite-basis all-electron approaches where most previous HL-MGGA developments have occurred.

The remainder of this paper is organized as follows. Sec.~\ref{sec:theory} introduces the general functional form of an HL-MGGA and outlines the construction principles and design strategy of $\vartheta$-PBE. 
Sec.~\ref{sec:scf} describes the derivation and implementation of the self-consistent procedure. 
Evaluations spanning a diverse set of molecular, solid-state, and heterogeneous catalysis properties are presented in Sec.~\ref{sec:results}. Finally, Sec.~\ref{sec:conclusion} summarizes our findings and main conclusions.

%% file: theory.tex
\section{Theory}
\label{sec:theory}

HL-MGGA functionals can be written as the semilocal form
\begin{equation}
	\label{eq:ef}
	\functional{E_{xc}}{n} = \int{\rd\vrr e_{xc}({n, {\nabla{n}}, {\nabla^{(2)}{n}}})}
	\equiv \int{\rd\vrr e_{xc}({{n}, {\nabla{n}}, f})},
\end{equation}
where $e_{xc}$ is the xc energy density and
the dependence on $\nabla^{(2)}n$ enters only through a scalar quantity
$f \equiv f(n, \nabla n, \nabla^{(2)}n)$, e.g. a function of Eq.~\eqref{eq:theta} or Eq.~\eqref{eq:rdh}. While the form of $f$ is in general arbitrary, in this work we restrict $f$ to be a $\vartheta$-dependent smooth switching function, i.e. $f(\vartheta)\in(0,1]$ of the form
\begin{equation}
    \label{eq:ftheta}
    \function{f}{\vartheta} = \frac{1}{1 + a\vartheta^2},
\end{equation}
so that $f(0)=1$ in single-orbital and asymptotic tail regions (single exponential densities), 
and $f(\vartheta\!\to\!\infty)\to\!0$ near bond critical points and slowly-varying density regions.
The scaling parameter $a > 0$ sets the transition scale between these two limits and 
was arbitrarily set to $a=1$ in $\vartheta$-MGGA,
but later in this section we will fix it to a larger value based on a physical constraint.

Following the $\vartheta$-MGGA approach, we construct the exchange-correlation (xc) functional as an interpolation between different parameterizations of PBE~\cite{pbe1996}, 
utilizing the exchange enhancement factor
\begin{equation}
  F_{x}(s) = 1 + \kappa - \frac{\kappa}{1 + \mu s^2 / \kappa}
\end{equation}
where the constant $\kappa=0.804$ is chosen to satisfy the Lieb-Oxford bound~\cite{lieb,perdew1991electronic}. 
The gradient coefficient $\mu$ and its correlation counterpart $\beta$ (Eqs.~7 and 8 of Ref.~\cite{pbe1996})
embody the difficulty of simultaneously satisfying multiple exact constraints at the GGA level.
In the original PBE, the correlation parameter 
$\beta_{\mathrm{cGE}} = 0.066725$ satisfies the second-order gradient expansion (GE) for correlation \cite{mabruckner},
while the exchange coefficient
$\mu_{\mathrm{cGE}} = 0.21951$
is chosen to recover the LDA linear response of the uniform-electron-gas,
as required by
\begin{equation}
  \mu = \frac{\pi^2}{3}\,\beta.
\end{equation}
However, this choice violates the second-order GE for exchange~\cite{ge2},
where $\mu_{\mathrm{xGE}} = 10/81 \approx 0.12346$.
The PBEsol functional~\cite{pbesol} improves the description of solids
by restoring $\mu = \mu_{\mathrm{xGE}}$ and fitting the correlation parameter to
$\beta = 0.046$ based on TPSS jellium xc energies.
Furthermore, an equivalent linear-response value of $\beta_{\mathrm{xGE}}\approx0.0375$ has been shown to yield competitive lattice constants~\cite{pbeparams},
corresponding to the $\beta(r_s \to \infty)$ limit of the density-dependent form used in revTPSS~\cite{revtpss}.

At the opposite extreme, the PBEmol functional~\cite{pbemol} improves molecular energetics by recovering the exact exchange energy of the hydrogen atom, 
utilizing $\mu_{xH} = 0.27853$ and the corresponding $\beta_{xH} \approx 0.08384$. 
Notably, this exact constraint is generally satisfied by meta-GGAs through a multiplicative factor in the $\tau = \tau^W$ limit~\cite{tpss2003,scan2015}.

Using the switching function of Eq.~\eqref{eq:ftheta}, we define the local gradient coefficient
for exchange as
\begin{equation}
  \label{eq:muint}
  \mu(\vartheta) = f(\vartheta)\,\mu_{xH} + \big[1 - f(\vartheta)\big]\,\mu_{xGE},
\end{equation}
and for correlation
\begin{equation}
  \label{eq:betaint}
  \beta(\vartheta) = f(\vartheta)\,\beta_{xH} + \big[1 - f(\vartheta)\big]\,\beta_{xGE}.
\end{equation}

\begin{figure}[tbp]
  \centering
  \includegraphics[width=0.45\textwidth]{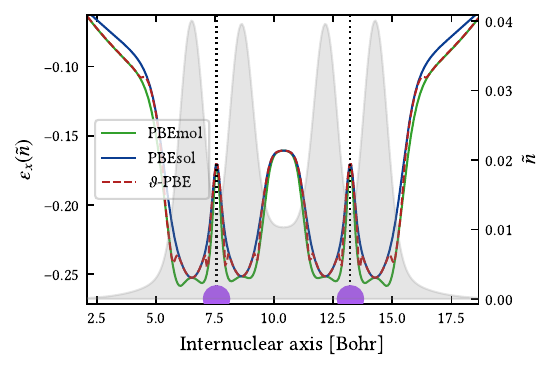}
  \caption{
    Exchange energy per particle along the internuclear axis of the Na$_2$ dimer. 
    The proposed $\vartheta$-PBE functional (dashed red line) interpolates smoothly between its parent functionals, 
    PBEmol (solid green line) and PBEsol (solid blue line). The gray background shading denotes the pseudo-density.
  }
  \label{fig:na2}
\end{figure}

This construction defines the $\vartheta$-PBE functional. 
By employing a purely PBE-like correlation rather than the modified TPSS correlation used in $\vartheta$-MGGA, 
we avoid numerical difficulties during self-consistent evaluations, albeit at the cost of losing the 
one-electron correlation-free property.
The interpolation between these limiting behaviors is demonstrated for a Na$_2$ dimer in Figure~\ref{fig:na2}, 
showing the exchange energy per particle ($\epsilon_x = e_x / n$). 
By design, $\vartheta$-PBE asymptotically approaches the PBEmol limit in the density tails. 
Crucially, in the bonding region between the atoms, $\vartheta$-PBE tracks PBEsol 
before reaching the uniform-electron-gas limit at the bond critical point where the density gradient vanishes.

\begin{figure*}[tbp]
  \centering
  \includegraphics[width=\textwidth]{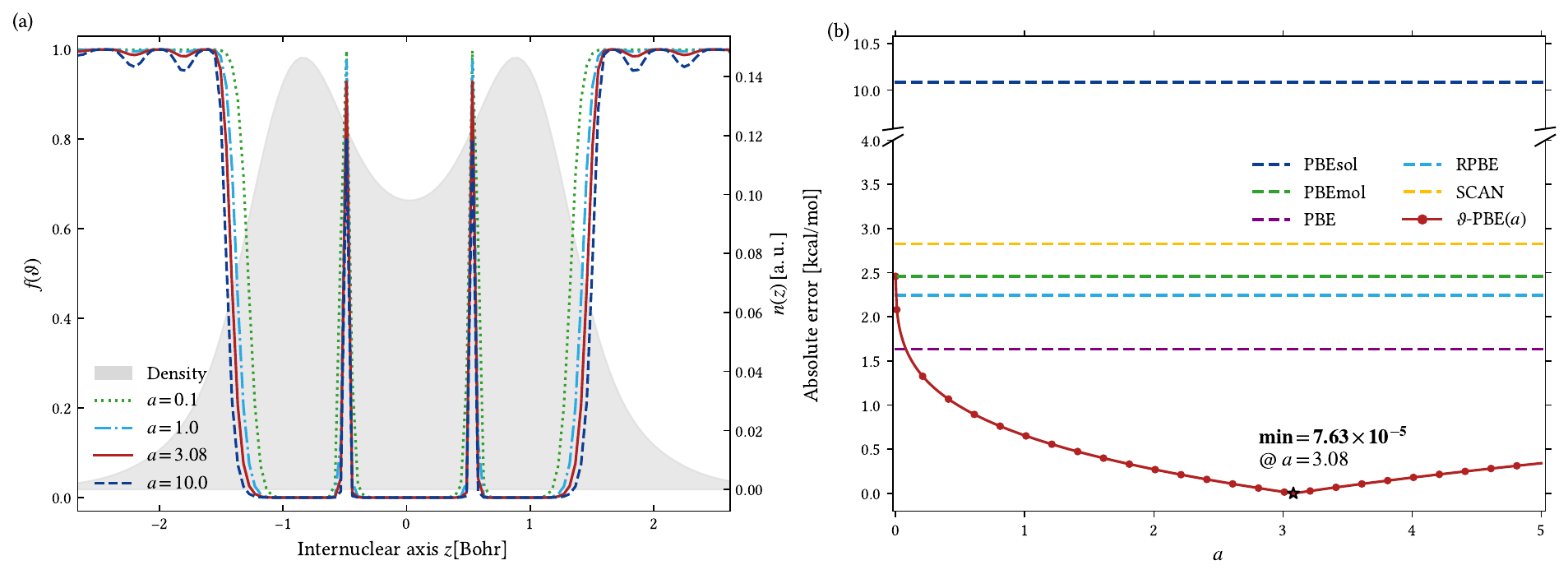}
  \caption{
  Fitting of the scaling parameter $a$ using the H$_2^+$ Hartree-Fock reference. 
  (a) Profile of the switching function $f(\vartheta)$ along the internuclear axis $z$ for $a = 0.1$, $1$, $3.08$, and $10$. 
  The shaded gray region represents the Hartree-Fock pseudo-density (right axis). 
  (b) Absolute error in the exchange energy (kcal/mol) of $\vartheta$-PBE as a function of $a$, calculated relative to the exact exchange energy of H$_2^+$ ($E_{\mathrm{exx}} \approx -206.46$ kcal/mol $\approx -0.329$ Ha). 
  The optimal parameter $a = 3.08$ (marked by the black star) minimizes this deviation. 
  Horizontal dashed lines indicate the absolute exchange energy errors of other standard semilocal functionals for comparison.
  }
  \label{fig:h2p}
\end{figure*}

For spin-polarized systems, the exchange energy obeys the exact spin-scaling relation~\cite{spinscaling-perdew} 
because $\vartheta$ is dimensionless and scale-invariant with respect to density.
For the correlation part, we employ a density-weighted mixing:
\begin{equation}
  f(\vartheta)
  = \sum_{\sigma=\uparrow,\downarrow}
    \frac{n_\sigma}{n}\,
    f\!\big(\vartheta_\sigma\big),
\end{equation}
where $n_\sigma$ and $\vartheta_\sigma$ denote the spin densities and their corresponding indicators.

To establish a meaningful interpolation between the two asymptotic regimes, we introduce an intermediate reference based on the simplest molecule, 
the dihydrogen cation ${\rm H}_2^+$, 
which is a prototypical example of the self-interaction error in DFT~\cite{h2plus-sie}.
The scaling parameter $a$ in Eq.~\ref{eq:ftheta} is therefore determined by minimizing the deviation between the 
$\vartheta$-PBE exchange energy and the exact exchange energy at the equilibrium bond length of 
H$_2^+$ ($R \approx 2\;\text{Bohr}$). 
Figure~\ref{fig:h2p}a demonstrates that varying $a$ 
preserves the overall shape of $\function{f}{\vartheta}$. However this variation systematically shifts the energies as illustrated in 
Figure~\ref{fig:h2p}b, which plots the corresponding exchange energy errors 
evaluated on the Hartree-Fock self-consistent density. 
Even without scaling, $\vartheta$-PBE yields a smaller error than its underlying GGA forms. 
Interestingly, PBEmol exhibits a larger error than PBE and RPBE \cite{rpbe1999}, a result mirrored by the SCAN functional. 
Since both functionals are explicitly constrained to satisfy the exact exchange of the isolated hydrogen atom, 
these increased deviations demonstrate
that a fixed single-electron limit ($\tau = \tau^W$) cannot simultaneously accommodate the energetics of both the atom and the H$_2^+$ bond.
In contrast, this limitation does not affect $\vartheta$-PBE as $\vartheta$ distinguishes between a single-exponential atomic density 
and the overlap region of the bond.
Minimization yields an optimal scaling parameter of $a = 3.08$. 
However, this fitting procedure does not entirely eliminate the one-electron self-interaction error of $\vartheta$-PBE, because the correlation energy does not vanish. 
Furthermore, it fails to reproduce the challenging dissociation curve of H$_2^+$ \cite{cohen2012challenges}, as the parameterization is performed solely at the equilibrium bond length.

The construction above only violates the second-order GE for correlation in PBE, which is known to matter less for real systems \cite{pbesol}, 
while enforcing the second-order GE for exchange
and the exact hydrogen exchange energy. 
Moreover, the choice of the scaling parameter $a$ can be viewed as satisfying an 
additional exact constraint or an appropriate norm that is not accessible to region indicators based on $\tau^W$.

%% file: scfimp.tex
\section{Self-consistent implementation}
\label{sec:scf}
We implemented the functional self-consistently in the GPAW code~\cite{gpaw2024,thetapbe} under the identifier \texttt{ThetaPBE} by constructing a general HL-MGGA interface.

The explicit density dependence of HL-MGGAs allows their straightforward implementation within existing GGA or LL-MGGA frameworks, without requiring the GKS formalism.
The key step in a self-consistent implementation is evaluating the xc potential as the functional derivative of the xc energy functional.
For a HL-MGGA of the form~\eqref{eq:ef}, the derivative follows directly from the second-order Euler-Lagrange equation,
\begin{equation}
\label{eq:vxc1}
  \funcr{v_{xc}}
  = \funcd{e_{xc}}{n}
  - \nabla\!\cdot\!\funcd{e_{xc}}{\nabla n}
  + \nabla^{(2)}\!\cdot\!\funcd{e_{xc}}{\nabla^{(2)} n},
\end{equation}
where the last term denotes a tensor contraction.
As mentioned in Sec.~\ref{sec:theory}, for $\vartheta$-PBE
the Hessian dependence enters through the dimensionless scalar
$f(n,\nabla n,\nabla^{(2)}n)$ which, by the chain rule, leads to
\begin{equation}
\label{eq:vxc2}
\begin{aligned}
    \funcr{v_{xc}}
  &= \funcr{v_{xc}^{\mathrm{GGA}}}
  + \funcd{e_{xc}}{f}\,\funcd{f}{n}
  - \nabla\!\cdot\!\left(\funcd{e_{xc}}{f}\,\funcd{f}{\nabla n}\right)
  \\
  &+ \nabla^{(2)}\!\cdot\!\left(\funcd{e_{xc}}{f}\,
  \hat S\!\left[\funcd{f}{\nabla^{(2)} n}\right]\!\right),
\end{aligned}
\end{equation}
where $\funcr{v_{xc}^{\mathrm{GGA}}}$ denotes the GGA-like part, and the symmetrization operator
$$
  \hat S[X] = X + X^{T} - \mathrm{diag}\,X,
$$
accounts for derivatives with respect to the symmetric Hessian~\cite{mcbook}.

To evaluate these quantities accurately across the entire simulation cell, we rely on the PAW method~\cite{blochl2003paw}. Conceptually, the PAW formalism provides a rigorous framework to reconstruct the highly oscillatory all-electron (AE) wavefunctions near the nuclei while solving the Kohn-Sham equations for smooth, computationally efficient pseudo (PS) wavefunctions in the interstitial regions. This is achieved by partitioning the system space: the interstitial region is treated using a suitable basis for smooth functions, whereas the regions close to the nuclei are enclosed in atomic augmentation spheres where quantities are expanded in terms of localized partial waves. 

Within this formalism, the semilocal xc energy is decomposed as
\begin{equation}
  \functional{E_{xc}}{n} = \functional{E_{xc}}{\tilde{n}} + 
  \sum_{a}{\left(
    \functional{E_{xc}}{n^{a}} - \functional{E_{xc}}{\tilde{n}^{a}}
\right)},
\end{equation}
where $\tilde{n}$ is the smooth PS density on the global grid plus a remainder pseudized core density, and $n^a$ and $\tilde n^a$ are the AE and PS on-site atomic densities represented on a radial grid inside the augmentation spheres \cite{pawvasp, paw2010, gpaw2024}.

Accordingly, two complementary strategies are dictated by the GPAW implementation for evaluating the required second derivatives. For the smooth PS density $\tilde{n}$, which is represented on a uniform real-space grid (or equivalently, via plane-waves), the gradients and Hessians are computed using high-order finite-difference stencils~\cite{fd-1994}. Inside the augmentation spheres, where the one-center quantities are represented in a spherical-harmonics expansion, gradients and Hessians are obtained through a combination of finite-differences on the radial grids and exact analytical evaluation of the angular parts. The atomic PAW corrections to the Hamiltonian are then evaluated using Eq.~\eqref{eq:vxc2}, where the terms involving partial derivatives of $f(\vartheta)$ with respect to $\nabla{n}$ and $\nabla^{(2)}n$ are handled by an integration-by-parts scheme (see Appendix~\ref{app:scfpaw}).

%% file: results.tex
\section{Results and Discussions}
\label{sec:results}
In this section, we evaluate $\vartheta$-PBE performance across a diverse range of physical and chemical properties to establish its general accuracy and limits of applicability. 
The benchmarks are divided into three distinct regimes. First, we test molecular properties 
to assess the functional's description of atomization energies and reaction energetics. 
Next, we examine bulk solid-state properties (lattice constants and cohesive energies) to probe its behavior in the slowly varying density limit characteristic of extended crystalline systems. 
Finally, we evaluate chemisorption energies, a regime where both localized molecular and delocalized solid-state electron densities are simultaneously present; 
capturing this balance accurately is a fundamental requirement for modeling heterogeneous catalysis.

All calculations were performed using a development version~\cite{thetapbe} of GPAW with standard PBE PAW setups, except for RPBE, for which consistent setups were employed. For Ni, a 10-valence-electron setup was utilized instead of the default 16-electron version to improve convergence.

\subsection{Molecular properties}
\label{subsec:resmol}
We first benchmark the performance of $\vartheta$-PBE against RPBE \cite{rpbe1999}, 
various PBE parameterizations, and the SCAN meta-GGA on molecular properties: atomization energy and barrier-height calculations for the AE6 and BH76 test sets \cite{ae6bh6,bh76}. 
The calculations were performed with a plane-wave cutoff of 1000 eV. Each system was placed at the center of a cubic box with at least 6 $\text{\AA}$ of vacuum in all directions, 
and slight cell perturbations were applied to break symmetry. 
Molecular geometries for AE6 were taken from Ref. \cite{ae6bh6ref}, and those for BH76 from the GMTKN55 database \cite{gmtkn55}. 

\begin{table}[tbp]
\caption{
Mean error (ME) and mean absolute error (MAE) for the AE6 (atomization energies) and BH76 (barrier heights) datasets. 
All values are in eV.
}
\label{tab:molbench}
\begin{ruledtabular}
\begin{tabular}{lccccccc}
    & & PBE & PBEsol & PBEmol & RPBE & SCAN & $\vartheta$-PBE \\
    \hline
    AE6  & ME  &  0.45 &  1.46 & -0.04 & -0.40 &  0.04 &  0.00 \\
         & MAE &  0.61 &  1.47 &  0.36 &  0.47 &  0.16 &  0.41 \\
    \hline
    BH76  & ME  & -0.39 & -0.46 & -0.33 & -0.29 & -0.33\footnote{\label{note:scanbh76}Reference~\cite{scan2015}} & -0.30 \\
         & MAE &  0.39 &  0.49 &  0.36 &  0.33 &  0.33\textsuperscript{\ref{note:scanbh76}} &  0.34 \\
\end{tabular}
\end{ruledtabular}
\end{table}

The AE6 dataset serves as a representative set for assessing the description of molecular bonds by comparing calculated atomization energies with accurate theoretical reference values \cite{ae6bh6ref}.
As summarized in Table \ref{tab:molbench}, $\vartheta$-PBE performs comparably to PBEmol on atomization energies, with an increase in the mean absolute error (MAE) of 0.05 eV. 
Neither functional reaches the accuracy of SCAN ($\text{MAE} = 0.16$ eV) on this benchmark. However, $\vartheta$-PBE eliminates the slight underbinding bias of PBEmol ($\text{ME}=-0.04$ eV); 
both functionals significantly improve upon the overbinding tendency of PBE ($\text{ME} = +0.45$ eV) and the systematic underbinding of RPBE ($\text{ME} = -0.40$ eV). 
As expected, the PBEsol functional performs poorly for molecular atomization ($\text{MAE} = 1.47$ eV).

For reaction barrier heights (BH76 dataset), $\vartheta$-PBE improves upon PBEmol by 0.02 eV and achieves an accuracy close to that of RPBE and SCAN. 
Although it still exhibits the systematic underestimation of barrier heights characteristic of semilocal functionals ($\text{ME} = -0.30$ eV), 
it reduces the error relative to PBE ($\text{ME} = -0.39$ eV). Notably, $\vartheta$-PBE succeeds in capturing the improved barrier description typically associated with RPBE, while simultaneously improving upon its atomization energies.

\subsection{Bulk properties}
\label{subsec:ressol}
To evaluate the performance of $\vartheta$-PBE on bulk properties, we calculated the lattice constants and cohesive energies for the LC20 dataset \cite{scfmgga_2011}, 
which provides zero-point-corrected experimental values for 20 cubic solids. 
The calculations used Brillouin-zone integrations with a $(17 \times 17 \times 17)$ Monkhorst-Pack grid \cite{monkhorst-pack},
a Fermi-Dirac smearing of 0.1 eV, and an 800 eV cutoff, except for Li systems (1200 eV for Li; 1000 eV for LiCl and LiF) following Refs. \cite{scfmgga_2011,scanl-bench}.
Equilibrium volumes were obtained from twelve-point energy-volume curves within $\pm 10\%$ of the minimum, 
fitted to the stabilized jellium equation of state (SJEOS) \cite{sjeos} as implemented in ASE \cite{ase-paper},
with stress converged below $5$ meV/$\text{\AA}^2$. To evaluate cohesive energies, isolated atoms were simulated in a $(10 \times 11 \times 12) \text{ \AA}^3$
box using spin-polarized calculations with the same plane-wave cutoff and fractional occupancies to maintain spherical symmetry.

\begin{figure*}[tbp]
  \centering
  \includegraphics[width=0.99\textwidth]{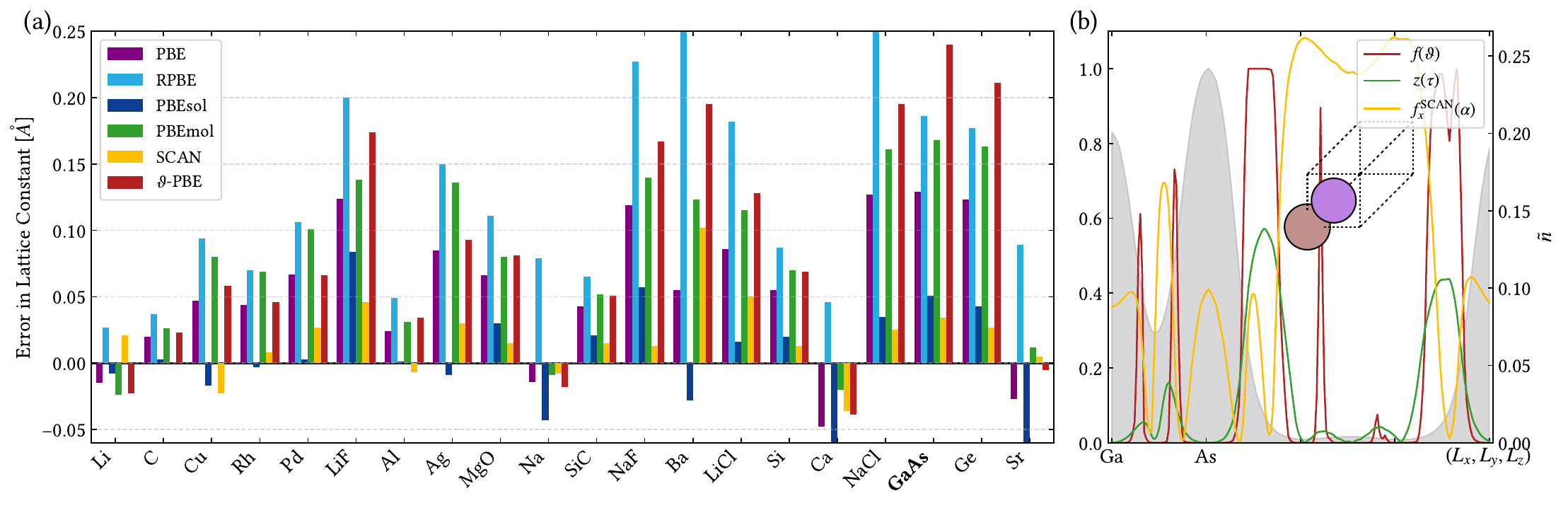}
  \caption{
    (a) Errors in calculated lattice-constants (in \AA) relative to experimental values for the LC20 database. 
    (b) GaAs region indicators in the PAW pseudo space. 
    The proposed switching function $f(\vartheta)$ (red) is compared against the iso-orbital indicator $z(\tau)$ (green) 
    and the SCAN switching function for exchange (yellow).
  }
  \label{fig:lc-gaas}
\end{figure*}

\begin{table}[tbp]
  \caption{ME and MAE for lattice constants ($a_0$ in \AA) and cohesive energies ($E_{\rm coh}$ in eV) calculated for the 20 solids in the LC20 database.}
\label{tab:solbench}
\begin{ruledtabular}
\begin{tabular}{lccccccc}
& & PBE & PBEsol & PBEmol & RPBE & SCAN &  $\vartheta$-PBE \\
\hline
$a_0$         &      ME  & 0.055  & 0.001  &  0.080  &  0.125 &   0.023  & 0.087  \\
              &      MAE & 0.066  & 0.036  &  0.086  &  0.125 &   0.025  & 0.096  \\
\hline
$E_{\rm coh}$ &      ME  & -0.09  & 0.24   &  -0.23  & -0.40  &  -0.10\footnote{\label{note:scanlc20}Reference~\cite{scanl-bench}}    &  -0.25 \\
              &      MAE & 0.14   & 0.26   &  0.23   &  0.40  &  0.24\textsuperscript{\ref{note:scanlc20}}    &  0.26  \\
\end{tabular}
\end{ruledtabular}
\end{table}
As summarized in Table \ref{tab:solbench}, $\vartheta$-PBE systematically overestimates lattice constants ($\text{ME} = 0.087 \text{ \AA}$, $\text{MAE} = 0.096 \text{ \AA}$), 
notably exceeding the error of PBEmol ($\text{MAE} = 0.086 \text{ \AA}$) despite the latter serving as the upper bound in the functional's interpolation scheme. 
This degradation likely stems from two interconnected sources. 
First, the simple interpolation in Eqs. \eqref{eq:muint} and \eqref{eq:betaint} using the continuous switching function $f(\vartheta)$ introduces a bias toward molecular properties, 
particularly for correlation, highlighting the necessity of the more complex mathematical constructions utilized in modern meta-GGAs. 
Second, the behavior of $\vartheta$ near the slowly-varying limit is mathematically fragile \cite{thetahyb2015}. 
For $\vartheta$ to correctly identify this limit, the squared reduced gradient $s^2$ must vanish more rapidly than the reduced density Hessian $p$ (Eq. \eqref{eq:theta_approx}). 
Consequently, $\vartheta$ sometimes fails to detect slowly-varying regions until $s^2$ is extremely small, at which point standard GGAs have already collapsed to the correct LDA limit. 
The error distribution across the LC20 set supports this mathematical rationale (Fig. \ref{fig:lc-gaas}a), with the largest deviations occurring in ionic systems and semiconductors, 
peaking at a $0.24 \text{ \AA}$ overestimation for GaAs. Figure \ref{fig:lc-gaas}b further illustrates the underlying cause by comparing the region indicator $f(\vartheta)$ 
against the iso-orbital indicator $z$ and the SCAN exchange switching function $f_x^{\mathrm{SCAN}}(\alpha)$ (Eq. 9 of Ref. \cite{scan2015}). 
While $f(\vartheta)$ and $z$ qualitatively agree on limiting regions, $f(\vartheta)$ transitions much more aggressively, and both differ significantly from $f_x^{\mathrm{SCAN}}(\alpha)$, 
particularly in low-density interstitial regions and atomic cores. This suggests that while the condition $|\nabla n| \to 0$ is necessary to detect the slowly-varying limit, 
it is insufficient for accurately describing intermediate regions without an additional anchor such as $\alpha$. 
Moving forward, incorporating concepts from differential geometry—specifically utilizing the shape operator \cite{shapeoperator} to measure the curvature of density isosurfaces—could 
provide a mathematically robust framework for designing Hessian-level region indicators that properly capture this limit.

Overall, the solid-state performance of $\vartheta$-PBE heavily mirrors PBEmol, preserving molecular accuracy at the
direct expense of bulk lattice predictions. 
For cohesive energies, $\vartheta$-PBE ($\text{MAE} = 0.26$ eV) performs comparably to PBEsol and PBEmol, but remains noticeably less accurate than standard PBE ($\text{MAE} = 0.14$ eV).

\subsection{Chemisorption energies}
Finally, we evaluate the performance of the functional for chemisorption energies, a highly relevant metric for heterogeneous catalysis. 
This is a regime where both density regions emphasized in the construction of $\vartheta$-PBE—molecular and slowly-varying—are simultaneously present. 
We exclude physisorption-dominated systems from this benchmark, as standard semilocal functionals inherently lack the necessary description of van der Waals dispersion interactions \cite{dftd,beef-vdw,araujo2022adsorption}. 

We employ the FCC subset of the chemisorption benchmark from Ref. \cite{ad_wellendorff2015}, consistent with the subset utilized in Ref. \cite{rtpss2018}. 
Chemisorption energies were evaluated using $(2 \times 2)$ four-layer FCC metal slabs separated by $10 \text{ \AA}$ of vacuum in both directions. 
The calculations assumed an adsorbate coverage of 1/4 ML and employed both self-consistent and fixed lattice constants. 
Following previous benchmarks \cite{rpbe1999,ad_wellendorff2015,rtpss2018,araujo2022adsorption,adsorption_validation}, all surface calculations used an 800 eV cutoff, $(8 \times 8 \times 1)$ 
Monkhorst-Pack grids, and second-order Methfessel-Paxton smearing \cite{methfessel-paxton} with a width of 0.2 eV. 
Ionic relaxations were performed using the BFGS algorithm until forces dropped below 0.05 eV/$\text{\AA}$ (0.08 eV/$\text{\AA}$ for Ni). 
For adsorption calculations, the lower two layers were fixed at bulk positions, whereas for the clean reference slabs, all atoms were allowed to relax. 
Adsorbates were initialized at atop, bridge, hollow, and fcc sites. Isolated reference energies were obtained in a $(13 \times 14 \times 15) \text{ \AA}^3$ supercell with full relaxation. 
Spin polarization was included for all Ni-containing systems and isolated open-shell adsorbates.

We first assess chemisorption energies using self-consistent lattice constants for the underlying bulk metals, the values of which are reported in Table~\ref{tab:chemisorption_lc},
comparing $\vartheta$-PBE directly to RPBE, a functional specifically designed to improve adsorption properties. 
The results for this self-consistent comparison are summarized in Table \ref{tab:chemisorption_lc} and visualized in Fig. \ref{fig:chemerr}. 
Overall, the two functionals deliver comparable accuracy: RPBE yields a smaller MAE (0.14 eV) with a positive mean error (indicating general underbinding), 
while $\vartheta$-PBE exhibits a marginally larger MAE (0.17 eV) with a negative mean error (indicating overbinding). 
Notably, $\vartheta$-PBE improves upon RPBE for all hydrogen adsorptions, whereas RPBE generally performs better for CO. 
For systems where RPBE shows relatively large deviations—such as O/Ni(111), NO/Pd(111), and CO/Pd(100)—$\vartheta$-PBE 
shifts the adsorption energy closer to the experimental reference. Conversely, for systems like CO/Ni(111) and O/Pt(111), the opposite trend is observed.

\begin{table}[tbp]
\caption{Calculated chemisorption energies (eV) for selected adsorbate/metal systems.
Results are obtained using lattice constants optimized self-consistently for each method.}
\label{tab:chemisorption_lc}
\centering
\begin{ruledtabular}
\begin{tabular}{lccc}
System & RPBE & $\vartheta$-PBE & Expt. \\
\hline
H/Pt(111)   & -2.59 &  -2.65    & -2.75 \\
H/Ni(100)   & -2.63 &  -2.72    & -2.82 \\
H/Ni(111)   & -2.67 &  -2.77    & -2.89 \\
H/Pd(111)   & -2.69 &  -2.70    & -2.84 \\
H/Rh(111)   & -2.69 &  -2.70    & -2.75 \\
O/Ni(100)   & -5.19 &  -5.54    & -5.36 \\
O/Ni(111)   & -4.93 &  -5.23    & -5.13 \\
O/Pt(111)   & -3.80 &  -3.95    & -3.70 \\
O/Rh(100)   & -4.66 &  -4.87    & -4.46 \\
I/Pt(111)   & -2.06 &  -2.17    & -2.40 \\
NO/Pd(111)  & -1.71 &  -1.89    & -1.89 \\
NO/Pd(100)  & -1.58 &  -1.77    & -1.69 \\
NO/Pt(111)  & -1.39 &  -1.51    & -1.23 \\
CO/Ni(111)  & -1.44 &  -1.79    & -1.28 \\
CO/Pd(111)  & -1.51 &  -1.69    & -1.49 \\
CO/Pd(100)  & -1.49 &  -1.65    & -1.63 \\
CO/Pt(111)  & -1.31 &  -1.45    & -1.28 \\
CO/Cu(111)  & -0.46 &  -0.54    & -0.59 \\
CO/Rh(111)  & -1.59 &  -1.73    & -1.47 \\
CO/Ir(111)  & -1.71 &  -1.83    & -1.70 \\
\hline
ME          & 0.06  &  -0.09    & -     \\
MAE         & 0.14  &   0.17    & -     \\
\end{tabular}
\end{ruledtabular}
\end{table}
\begin{table}[tbp]
\caption{
Calculated chemisorption energies (eV) for the selected adsorbate/metal systems within a fixed bulk lattice constant (SCAN equilibrium values) across all methods.
}
\label{tab:chemisorption_full}
\centering
\begin{ruledtabular}
\begin{tabular}{lccccccc}
     & PBE & PBEsol & PBEmol & RPBE & SCAN & $\vartheta$-PBE \\ \hline
    H/Pt(111)& -2.68 & -2.89 & -2.58 & -2.57 & -3.06 & -2.62 \\
    H/Ni(100)& -2.79 & -3.06 & -2.66 & -2.64 & -2.64 & -2.75 \\
    H/Ni(111)& -2.8 & -3.06 & -2.68 & -2.65 & -3.05 & -2.7 \\
    H/Pd(111)& -2.81 & -3.09 & -2.67 & -2.65 & -3.01 & -2.68 \\
    H/Rh(111)& -2.79 & -3.06 & -2.65 & -2.64 & -2.98 & -2.65 \\
    O/Ni(100)& -6.04 & -6.25 & -5.46 & -5.18 & -5.18 & -5.52 \\
    O/Ni(111)& -5.38 & -5.92 & -5.15 & -4.86 & -5.49 & -5.14 \\
    O/Pt(111)& -4.09 & -4.69 & -3.82 & -3.53 & -4.17 & -3.77 \\
    O/Rh(100)& -5.18 & -5.72 & -4.95 & -4.66 & -5.09 & -4.88 \\
    I/Pt(111)& -2.31 & -2.85 & -2.09 & -1.89 & -2.68 & -2.02 \\
    NO/Pd(111)& -2.09 & -2.64 & -1.85 & -1.62 & -2.3 & -1.85 \\
    NO/Pd(100)& -2.03 & -2.49 & -1.82 & -1.61 & -2.14 & -1.78 \\
    NO/Pt(111)& -1.68 & -2.23 & -1.44 & -1.23 & -1.92 & -1.41 \\
    CO/Ni(111)& -1.82 & -2.3 & -1.61 & -1.41 & -2.0 & -1.69 \\
    CO/Pd(111)& -1.86 & -2.36 & -1.63 & -1.44 & -2.13 & -1.65 \\
    CO/Pd(100)& -1.85 & -2.24 & -1.67 & -1.49 & -2.07 & -1.66 \\
    CO/Pt(111)& -1.47 & -1.89 & -1.28 & -1.13 & -1.9 & -1.3 \\
    CO/Cu(111)& -0.7 & -1.03 & -0.54 & -0.38 & -0.88 & -0.5 \\
    CO/Rh(111)& -1.88 & -2.22 & -1.72 & -1.58 & -2.12 & -1.7 \\
    CO/Ir(111)& -1.94 & -2.26 & -1.8 & -1.66 & -2.25 & -1.77 \\ \hline
    ME & -0.24 & -0.64 & -0.03 & 0.13 & -0.38 & -0.03 \\
    MAE & 0.27 & 0.64 & 0.16 & 0.17 & 0.42 & 0.15 \\
\end{tabular}
\end{ruledtabular}
\end{table}

To prevent fortuitous error cancellation between overestimated bulk volumes and surface energies, we performed a second benchmark where lattice constants were fixed to the SCAN equilibrium values, 
but the nuclei positions were allowed to fully relax. This disentangles the intrinsic electronic performance from bulk expansion artifacts, 
providing a direct comparison of surface energetics at a consistent lattice. As shown in Table \ref{tab:chemisorption_full}, 
standard PBE, PBEsol, and SCAN systematically overbind adsorbates, with particularly large errors for PBEsol and SCAN (MAE of 0.64 eV and 0.42 eV, respectively). 
RPBE remains a strong reference and is the only functional with a positive mean error (0.13 eV). Notably, PBEmol and $\vartheta$-PBE perform remarkably well: both exhibit virtually no systematic bias ($\text{ME} \approx -0.03$ eV) and achieve low MAEs of 0.16 eV and 0.15 eV, respectively. 

While the strong performance of $\vartheta$-PBE in this regime broadly mirrors that of PBEmol, it is not merely a reproduction. 
For specific systems like H/Ni(100) or CO/Ni(111), it diverges significantly, providing distinct physical corrections. 
We attribute this baseline success to the constraint of satisfying the exact exchange energy of the hydrogen atom. 
Interestingly, the well-known accuracy of RPBE for chemisorption can be traced to a similar mathematical feature. 
Its exponential exchange enhancement factor (Eq. 15 of Ref. \cite{rpbe1999}) with the standard choice of $\mu=\mu_{cGE}$ nearly reproduces the exact hydrogen exchange energy. 
If one performs a PBEmol-like refitting of $\mu$ for RPBE, the resulting parameter $\mu^{\mathrm{RPBEmol}} \approx 0.2283$ is about 4\% larger than $\mu_{cGE}$. 
This suggests that RPBE is already close to the molecularly optimal regime. Consequently, $\vartheta$-PBE provides chemisorption energies on par with RPBE while avoiding a strong systematic bias. 
Combined with its accurate atomization energies and barrier heights predictions, this makes $\vartheta$-PBE a promising choice for catalytic and surface-science applications, 
with the caveat that its equilibrium bulk volumes are generally not reliable.

%% file: summary.tex
\section{Conclusions}
\label{sec:conclusion}
In summary, we have introduced a self-consistent implementation of Hessian-level meta-GGA functionals within the PAW formalism. To achieve this, we overcame the primary technical challenges associated with higher-order density derivatives: deriving the exact analytical functional derivatives of the exchange-correlation energy with respect to the full density Hessian matrix, and formally extending the PAW corrections to handle Hessian-level corrections by utilizing the divergence theorem. Our implementation demonstrates the feasibility of utilizing higher-order density derivatives in practical DFT calculations and enables numerically stable calculations for both molecular and periodic systems.

As a prototypical HL-MGGA we introduced and benchmarked $\vartheta$-PBE, which is a modification of a previously proposed $\vartheta$-MGGA. Both functionals are constructed by interpolating between distinct PBE parameterizations, but the latter was implemented only post-SCF in an all-electron molecular code based on Slater basis functions. While $\vartheta$-PBE yields accurate chemisorption energies, it does not provide a uniform improvement over its underlying GGAs across all metrics---particularly 
regarding bulk lattice constants---and entails a higher computational cost due to the evaluation of the density Hessian. We have identified the main sources of the functional's overall unsatisfactory performance for materials. While the constructed switching function recovers the correct limits, in practice it biases the functional too much toward the molecular density regime in energetically relevant regions. Nevertheless, we have demonstrated that the HL-MGGA framework offers a systematic pathway for designing non-empirical, orbital independent functionals that utilizes the full density Hessian, going beyond standard Laplacian-level approximations.
Crucially, our results show that the variable $\vartheta$ successfully distinguishes between distinct one-electron density limits, 
specifically the single-center atomic density versus the two-center bond in H$_2^+$, a topological distinction that standard iso-orbital indicators often miss. 
This work therefore serves as a proof of principle that the complete set of second-order density derivatives can be effectively 
harnessed to construct non-empirical exchange-correlation functionals without sacrificing universality.
\section*{Acknowledgments}
This work was supported by the Novo Nordisk Foundation Data Science Research Infrastructure 2022 Grant: A high-performance computing infrastructure for data-driven research on sustainable energy materials, Grant no. NNF22OC0078009. We thank Jens Jørgen Mortensen for fruitful discussions and guidance on GPAW.
\section*{Code and data availability}
The implementation is publicly available via a merge request in the GPAW repository~\cite{thetapbe}.

%% file: appendix.tex
\appendix
\setcounter{table}{0}
\setcounter{figure}{0}
\renewcommand{\thetable}{\thesection.\arabic{table}}
\renewcommand{\thefigure}{\thesection.\arabic{figure}}
\section{Self-consistent implementation in PAW}
\label{app:scfpaw}

In the PAW method, the mapping from the PS to the AE space is defined through a linear transformation operator,
\begin{equation}
    \label{eq:transformation}
    \hat{T} \altket{\tilde{\psi}} = \altket{\psi}, \qquad
    \hat{T} = 1 + \sum_a \sum_i 
    \big(\altket{\phi_i^a} - \altket{\tilde{\phi}_i^a}\big)
    \altbra{\tilde{p}_i^a},
\end{equation}
where $\funcr{\tilde{p}_i^a}$ are the projector functions, 
$\funcr{\phi_i^a}$ the AE partial waves, and 
$\funcr{\tilde{\phi}_i^a}$ the corresponding PS partial waves. 
Outside each augmentation sphere of radius $r_c^a$, these partial waves are identical,
\begin{equation}
	\label{eq:partial-waves}
    \funcr{\phi_i^a} = \funcr{\tilde{\phi}_i^a}, 
    \qquad r > r_c^a.
\end{equation}
Typically, the smooth wavefunctions are expanded in terms of plane-waves while inside each augmentation sphere the partial waves $\funcr{\phi_i^a}$ and $\funcr{\tilde{\phi}_i^a}$ are expanded as products of radial functions and spherical harmonics.

The Kohn-Sham equations are therefore solved in the transformed (PS) representation as
\begin{equation}
    \label{eq:kst}
    \hat{T}^\dagger\, \functional{\hat{H}}{n}\, \hat{T} \altket{\tilde{\psi}_n}
    = \epsilon_n\, \hat{T}^\dagger \hat{T}\, \altket{\tilde{\psi}_n}.
\end{equation}
Substituting Eq.~\eqref{eq:transformation} leads to the PAW Hamiltonian,
\begin{equation}
    \hat{\tilde{H}} 
    = -\frac{1}{2}\nabla^2 
    + \funcr{\tilde{v}_{\mathrm{KS}}}
    + \sum_a \sum_{ii'} 
      \altket{\tilde{p}_i^a}\,
      \Delta H^a_{ii'}\,
      \altbra{\tilde{p}_{i'}^a},
\end{equation}
where $\funcr{\tilde{v}_{\mathrm{KS}}}$ is the smooth effective PS potential and 
$\Delta H^a_{ii'}$ the on-site PAW correction.

The total electron density in PAW is decomposed as
\begin{equation}
\label{eq:paw-decomp}
  n(\mathbf r) = \tilde n(\mathbf r)
  + \sum_a \big(n^a(\mathbf r) - \tilde n^a(\mathbf r)\big),
\end{equation}
where the smooth pseudo-density is
\[
  \tilde{n} = \sum_n f_n |\funcr{\tilde{\psi}_n}|^2 + \funcr{\tilde{n}_c},
\]
and $n^a$ and $\tilde n^a$ denote the AE and PS on-site atomic densities within the augmentation region $\Omega_a$.  
These are constructed from the atomic density matrices,
\begin{equation}
    \begin{aligned}
      n_\sigma^a(\mathbf r)
      &= \sum_{ii'} D^a_{\sigma ii'}\,
      \phi_i^a(\mathbf r)\,\phi_{i'}^a(\mathbf r)
      + \funcr{n^a_c}, \\
      D^a_{\sigma ii'} 
      &= \sum_n f_n
      \altbraket{\tilde\psi_{\sigma n}}{\tilde p_i^a}
      \altbraket{\tilde p_{i'}^a}{\tilde\psi_{\sigma n}},
    \end{aligned}
\end{equation}
and analogously for $\tilde n_\sigma^a$ using $\tilde{\phi}_i^a$.

The exchange-correlation (xc) potential contributes to the Hamiltonian through a PS part,
\begin{equation}
	\funcr{\tilde{v}_{xc}} 
  = \functionald{\functional{E_{xc}}{\tilde{n}}}{\tilde{n}},
\end{equation}
where $\funcr{\tilde{v}_{xc}}$ is obtained directly from Eq.~\eqref{eq:vxc2},
and through an on-site AE-PS correction,
\begin{equation}
    \label{eq:deltaH}
    \begin{aligned}
	\Delta H^a_{xc,i_1i_2} 
  &= \int\!\rd\vrr\, 
     \phi_{i_1}^a(\mathbf r)
     \functional{v_{xc}^a}{n^a}\,
     \phi_{i_2}^a(\mathbf r) \\
	 &\quad- 
     \int\!\rd\vrr\,
     \tilde{\phi}_{i_1}^a(\mathbf r)
     \functional{v_{xc}^a}{\tilde{n}^a}\,
     \tilde{\phi}_{i_2}^a(\mathbf r).
    \end{aligned}
\end{equation}
Here, $\phi_i^a$ and $\tilde{\phi}_i^a$ are the corresponding AE and PS partial waves inside $\Omega_a$.

To evaluate the xc contributions involving gradients and Hessians, 
we apply the divergence theorem within each PAW sphere $\Omega_a$, with $\hat{\mathbf r}_a$ denoting the outward unit normal on its surface $\partial\Omega_a$:
\begin{equation}
\begin{aligned}
\int_{\Omega_a}\!\rd\vrr\,
  &\big(\nabla\!\cdot\!\funcr{\mathcal{A}}\big)\,
  \phi^a_{i_1}\phi^a_{i_2}
= 
-\!\int_{\Omega_a}\!\rd\vrr\,
  \funcr{\mathcal{A}}\!\cdot\!
  \nabla\!\big(\phi^a_{i_1}\phi^a_{i_2}\big) \\
&+\oint_{\partial\Omega_a}\!\rd S\,
  (\funcr{\mathcal{A}}\!\cdot\!\hat{\mathbf r}_a)\,
  \phi^a_{i_1}\phi^a_{i_2},
\end{aligned}
\end{equation}
and
\begin{equation}
\begin{aligned}
\int_{\Omega_a}\!\rd\vrr\,
&\big(\nabla^{(2)}\!\cdot\!\funcr{\mathcal{B}}\big)\,
  \phi^a_{i_1}\phi^a_{i_2}
= 
\int_{\Omega_a}\!\rd\vrr\,
\funcr{\mathcal{B}}\!\cdot\!
\nabla^{(2)}\!\big(\phi^a_{i_1}\phi^a_{i_2}\big) \\ 
&+\oint_{\partial\Omega_a}\!\rd S\,
  \Big[
    \phi^a_{i_1}\phi^a_{i_2}\,
    \hat{\mathbf r}_a\!\cdot\!
    \big(\nabla\!\cdot\!\funcr{\mathcal{B}}\big)
    - 
    \nabla\!\big(\phi^a_{i_1}\phi^a_{i_2}\big)\!\cdot\!
    \big(\funcr{\mathcal{B}}\,\hat{\mathbf r}_a\big)
  \Big],
\end{aligned}
\end{equation}
for the AE part, where
\[
\funcr{\mathcal{A}} = 
\funcd{e_{xc}}{f}\,
\funcd{f}{\nabla n^a},
\qquad
\funcr{\mathcal{B}} = 
\funcd{e_{xc}}{f}\,
\funcd{f}{\nabla^{(2)} n^a}
\]
are the vector and tensor fields corresponding to the gradient and Hessian contributions, respectively.  
The same expressions apply for the PS quantities, and substitution into Eq.~\eqref{eq:deltaH} shows that the surface integrals cancel in the AE-PS difference for $\Delta H^a_{xc,ii'}$, 
as the PAW formalism guarantees continuity across $\partial\Omega_a$.

\section{Partial derivatives of \texorpdfstring{$\vartheta$}{θ}}
\label{app:dtheta}

The derivative of the switching function $f(\vartheta)$ [Eq.~\eqref{eq:ftheta}] with respect to $\vartheta$ is
\begin{equation}
    \funcd{\funcr{f}}{\vartheta}
    = \frac{-2a\vartheta}{\big(1 + a\vartheta^2\big)^2}.
\end{equation}

\paragraph*{Derivative with respect to the density.}
\begin{equation}
    \funcd{\vartheta}{n}
    = -4\,\frac{\alpha_1}{\gamma^2}
      + 2\,\frac{n\alpha_2}{\gamma^3}.
\end{equation}
with $\gamma = \nabla^T{n}\nabla{n}$, $\alpha_1 = \nabla^T{n}\nabla{\gamma}$, and $\alpha_2 = \nabla^T{\gamma}\nabla{\gamma}$.

\paragraph*{Derivative with respect to the density gradient.}
\begin{equation}
\label{eq:dthetadgrho1}
\begin{aligned}
    \funcd{\vartheta}{\gamma} &= 8\,\frac{n}{\gamma^3}\alpha_1
    - 3\,\frac{n^2}{\gamma^4}\alpha_2, \\
    \funcd{\vartheta}{\alpha_1} &= -4\,\frac{n}{\gamma^2}, \qquad
    \funcd{\vartheta}{\alpha_2} = \frac{n^2}{\gamma^3}.
\end{aligned}
\end{equation}

The auxiliary derivatives are
\begin{equation}
\label{eq:dthetadghrho2}
\begin{aligned}
    \funcd{\alpha_1}{\big(\nrho\big)} &=
    2\big(\hessian{n} + \hessian{n}^T\big)\nrho
    = 4\,\hessian{n}\nrho
    = 2\,\nabla\gamma, \\
    \funcd{\alpha_2}{\big(\nrho\big)} &=
    4\big(\hessian{n}^T\hessian{n} + \hessian{n}\hessian{n}^T\big)\nrho
    = 8\,\hessian{n}^2\nrho
    = 4\,\hessian{n}\,\nabla\gamma.
\end{aligned}
\end{equation}

Combining these gives
\begin{equation}
\label{eq:dthetadgrho}
\begin{aligned}
    \funcd{\vartheta}{\big(\nrho\big)} 
    &= \funcd{\vartheta}{\gamma}\funcd{\gamma}{(\nrho)}
     + \funcd{\vartheta}{\alpha_1}\funcd{\alpha_1}{(\nrho)}
     + \funcd{\vartheta}{\alpha_2}\funcd{\alpha_2}{(\nrho)} \\
    &= 2\!\left(8\frac{n}{\gamma^3}\alpha_1
      - 3\frac{n^2}{\gamma^4}\alpha_2\right)\!\nrho
     - 4\frac{n}{\gamma^2}(4\hessian{n}\nrho)
     + \frac{n^2}{\gamma^3}(8\hessian{n}^2\nrho) \\
    &= \frac{2n}{\gamma^3}
      \Big(8\alpha_1 - \frac{3n\alpha_2}{\gamma}
      - 8\gamma\hessian{n} + 4n\hessian{n}^2\Big)\nrho.
\end{aligned}
\end{equation}

An equivalent and more compact form is
\begin{equation}
\label{eq:dthetadgrho3}
\begin{aligned}
    \funcd{\vartheta}{\big(\nrho\big)} &=
    2\funcd{\vartheta}{\alpha_1}\!\left(
    \nabla\gamma - \frac{2}{\gamma}\alpha_1\nrho\right)
    + 2\funcd{\vartheta}{\alpha_2}\!\left(
    2\hessian{n}\nabla\gamma - \frac{3}{\gamma}\alpha_2\nrho\right).
\end{aligned}
\end{equation}

\paragraph*{Derivative with respect to the gradient of the squared density gradient.}
\begin{equation}
\label{eq:dthetadgsigma}
\begin{aligned}
    \funcd{\vartheta}{\big(\nabla\gamma\big)} 
    &= -4\,\frac{n}{\gamma^2}
       \funcd{\alpha_1}{\big(\nabla\gamma\big)}
       + \frac{n^2}{\gamma^3}
         \funcd{\alpha_2}{\big(\nabla\gamma\big)}, \\
    \funcd{\alpha_1}{\big(\nabla\gamma\big)} &= \nrho, \\
    \funcd{\alpha_2}{\big(\nabla\gamma\big)} &= 2\,\nabla\gamma.
\end{aligned}
\end{equation}

\paragraph*{Derivative with respect to the density Hessian.}
\begin{equation}
\label{eq:dthetadH}
\begin{aligned}
    \funcd{\vartheta}{\hessian{n}}
    &= \funcd{\vartheta}{(\nabla\gamma)}
       \funcd{(\nabla\gamma)}{\hessian{n}}
     = 2\,\funcd{\vartheta}{(\nabla\gamma)}\,\nrhot.
\end{aligned}
\end{equation}

Alternatively, one may take the derivative directly:
\begin{equation}
    \funcd{\vartheta}{\hessian{n}}
    = -8\,\frac{n}{\gamma^2}\nrho\nrhot
      + 8\,\frac{n^2}{\gamma^3}
        \big(\hessian{n}\nrho\nrhot\big).
\end{equation}

Since the Hessian matrix is symmetric, the derivative with respect to the structured matrix is symmetrized as
\begin{equation}
    \funcdif{\vartheta}{\hessian{n}}
    = \big[\funcd{\vartheta}{\hessian{n}}\big]
      + \big[\funcd{\vartheta}{\hessian{n}}\big]^{\!T}
      - \mathrm{diag}\!\big[\funcd{\vartheta}{\hessian{n}}\big].
\end{equation}

\section{Calculations}
\label{app:calc}

\begin{table}[tbp]
  \caption{Equilibrium lattice constants, $a_0$ (\AA) for the LC20 solids.}
  \label{tab:lc20-results}
    \centering
    \begin{ruledtabular}
    \begin{tabular}{lccccccc}
        Solids & PBE & PBEsol & PBEmol & RPBE & SCAN & $\vartheta$-PBE & Expt. \\ \hline 
        Li & 3.438 & 3.445 & 3.429 & 3.480 & 3.474 & 3.430 & 3.453 \\ 
        Na & 4.200 & 4.171 & 4.205 & 4.293 & 4.206 & 4.196 & 4.214 \\ 
        Ca & 5.505 & 5.434 & 5.533 & 5.599 & 5.517 & 5.514 & 5.553 \\ 
        Sr & 6.018 & 5.923 & 6.057 & 6.134 & 6.050 & 6.040 & 6.045 \\ 
        Ba & 5.050 & 4.967 & 5.118 & 5.247 & 5.097 & 5.190 & 4.995 \\ 
        Al & 4.042 & 4.019 & 4.049 & 4.067 & 4.011 & 4.052 & 4.018 \\ 
        Cu & 3.642 & 3.578 & 3.675 & 3.689 & 3.572 & 3.653 & 3.595 \\ 
        Rh & 3.838 & 3.791 & 3.863 & 3.864 & 3.802 & 3.840 & 3.794 \\ 
        Pd & 3.943 & 3.879 & 3.977 & 3.982 & 3.903 & 3.942 & 3.876 \\ 
        Ag & 4.147 & 4.053 & 4.198 & 4.212 & 4.092 & 4.155 & 4.062 \\ 
        C & 3.573 & 3.556 & 3.579 & 3.590 & 3.553 & 3.576 & 3.553 \\  
        SiC & 4.389 & 4.367 & 4.398 & 4.411 & 4.361 & 4.397 & 4.346 \\ 
        Si & 5.476 & 5.441 & 5.491 & 5.508 & 5.434 & 5.490 & 5.421 \\ 
        Ge & 5.767 & 5.687 & 5.807 & 5.821 & 5.671 & 5.855 & 5.644 \\ 
        GaAs & 5.769 & 5.691 & 5.808 & 5.826 & 5.674 & 5.880 & 5.64 \\ 
        LiF & 4.096 & 4.056 & 4.110 & 4.172 & 4.018 & 4.146 & 3.972 \\ 
        LiCl & 5.156 & 5.086 & 5.185 & 5.252 & 5.120 & 5.198 & 5.07 \\ 
        NaF & 4.701 & 4.639 & 4.722 & 4.809 & 4.595 & 4.749 & 4.582 \\ 
        NaCl & 5.696 & 5.604 & 5.730 & 5.840 & 5.594 & 5.764 & 5.569 \\ 
        MgO & 4.255 & 4.219 & 4.269 & 4.300 & 4.204 & 4.270 & 4.189 \\ \hline 
        ME & 0.055 & 0.001 & 0.080 & 0.125 & 0.018 & 0.087 & -- \\ 
        MAE & 0.066 & 0.036 & 0.086 & 0.125 & 0.025 & 0.096 & -- \\ 
    \end{tabular}
\end{ruledtabular}
\end{table}

\begin{table}[tbp]
    \centering
    \caption{Cohesive energies, $E_{\rm coh}$ (eV/atom) of LC20 solids.}
    \begin{ruledtabular}
    \begin{tabular}{lccccccc}
        Solids & PBE &   PBEsol & PBEmol & RPBE & $\vartheta$-PBE & Expt. \\ \hline
        Ag &     2.53 &  3.09 &   2.29 &   2.02 & 2.23 &           2.98 \\ 
        Al &     3.5 &   3.85 &   3.36 &   3.25 & 3.27 &           3.43 \\ 
        Ba &     1.78 &  2.01 &   1.7 &    1.56 & 1.64 &           1.91 \\ 
        C &      7.82 &  8.32 &   7.59 &   7.33 & 7.65 &           7.54 \\ 
        Ca &     1.87 &  2.07 &   1.79 &   1.68 & 1.81 &           1.86 \\ 
        Cu &     3.47 &  4.01 &   3.23 &   2.98 & 3.18 &           3.52 \\ 
        GaAs &   3.18 &  3.55 &   3.01 &   2.8 &  2.85 &           3.34 \\ 
        Ge &     3.73 &  4.13 &   3.56 &   3.36 & 3.4 &            3.92 \\ 
        Li &     1.6 &   1.66 &   1.57 &   1.52 & 1.59 &           1.66 \\ 
        LiCl &   3.39 &  3.52 &   3.33 &   3.22 & 3.32 &           3.59 \\ 
        LiF &    4.42 &  4.51 &   4.38 &   4.23 & 4.37 &           4.46 \\ 
        MgO &    5.17 &  5.44 &   5.05 &   4.82 & 5.0 &            5.2 \\ 
        Na &     1.07 &  1.14 &   1.03 &   1.0 &  1.04 &           1.12 \\ 
        NaCl &   3.12 &  3.23 &   3.08 &   2.99 & 3.04 &           3.34 \\ 
        NaF &    3.9 &   3.99 &   3.87 &   3.74 & 3.82 &           3.97 \\ 
        Pd &     3.7 &   4.42 &   3.39 &   3.08 & 3.46 &           3.94 \\ 
        Rh &     5.96 &  6.81 &   5.58 &   5.29 & 5.7 &            5.78 \\ 
        Si &     4.56 &  4.91 &   4.4 &    4.25 & 4.32 &           4.68 \\ 
        SiC &    6.43 &  6.86 &   6.24 &   6.02 & 6.22 &           6.48 \\ 
        Sr &     1.57 &  1.77 &   1.49 &   1.39 & 1.49 &           1.73 \\ \hline
        ME &     -0.09 & 0.24 &   -0.23 &  -0.4 & -0.25 &          ~ \\
        MAE &    0.14 &  0.26 &   0.23 &   0.4 &  0.26 &           ~ \\
    \end{tabular}
    \end{ruledtabular}
\end{table}

\clearpage

\begingroup
\renewcommand{\arraystretch}{0.65} 
\setlength{\tabcolsep}{8pt}
\begin{longtable}{lccccccc}
  \caption{Calculated barrier heights from BH76 dataset. All values are in kcal/mol.} \\
  \hline
  System                         & PBE                       &PBEsol  & PBEmol & RPBE    &$\vartheta$-PBE  & Expt.\\ \hline
  \endfirsthead
  \multicolumn{8}{c}{{\bfseries \tablename\ \thetable{} -- continued from previous page}} \\
  \hline
  System                         & PBE                       & PBEsol & PBEmol & RPBE    & $\vartheta$-PBE & Expt. \\ \hline
  \endhead
  \hline
  \endfoot
h\_n2o\_n2ohts                   &9.9                        &7.4     &11.2    &11.5        &12.2            &17.7\\
oh\_n2\_n2ohts                   &49.0                       &41.0    &52.5    &55.8        &52.9            &82.6\\
h\_hf\_hfhts                     &26.4                       &23.1    &28.1    &28.7        &28.7            &42.1\\
hf\_h\_hfhts                     &26.4                       &23.1    &28.1    &28.7        &28.7            &42.1\\
h\_hcl\_hclhts                   &9.3                        &6.2     &11.0    &11.1        &11.3            &17.8\\
hcl\_h\_hclhts                   &9.3                        &6.2     &11.0    &11.1        &11.3            &17.8\\
h\_ch3f\_hfch3ts                 &18.0                       &16.6    &18.8    &19.0        &20.1            &30.5\\
hf\_ch3\_hfch3ts                 &40.3                       &36.8    &41.7    &44.5        &43.8            &56.9\\
h\_f2\_hf2ts                     &-9.6                       &-11.2   &-8.9    &-7.9        &-8.2            &1.5\\
hf\_f\_hf2ts                     &74.5                       &72.3    &75.1    &78.4        &78.1            &104.8\\
ch3\_clf\_ch3fclts               &-5.9                       &-7.0    &-5.6    &-3.6        &-3.8            &7.1\\
ch3f\_cl\_ch3fclts               &40.9                       &40.9    &40.7    &42.3        &43.2            &59.8\\
f-\_ch3f\_fch3fts                &0.2                        &1.1     &2.7     &5.0         &3.2             &-0.6\\
ch3f\_f-\_fch3fts                &0.2                        &1.1     &2.7     &5.0         &3.2             &-0.6\\
fch3fcomp\_fch3fts               &7.2                        &7.1     &7.5     &8.1         &8.7             &13.4\\
fch3fcomp\_fch3fts               &7.2                        &7.1     &7.5     &8.1         &8.7             &13.4\\
cl-\_ch3cl\_clch3clts            &3.6                        &4.9     &6.0     &7.7         &6.2             &2.5\\
ch3cl\_cl-\_clch3clts            &3.6                        &4.9     &6.0     &7.7         &6.2             &2.5\\
clch3clcomp\_clch3clts           &7.2                        &7.1     &7.3     &7.9         &8.7             &13.5\\
clch3clcomp\_clch3clts           &7.2                        &7.1     &7.3     &7.9         &8.7             &13.5\\
f-\_ch3cl\_fch3clts              &-9.5                       &-8.4    &-6.3    &-4.4        &-6.4            &-12.3\\
cl-\_ch3f\_fch3clts              &18.0                       &19.4    &19.6    &21.6        &20.0            &19.8\\
fch3clcomp1\_fch3clts            &-0.2                       &-0.4    &0.0     &0.2         &1.0             &3.5\\
fch3clcomp2\_fch3clts            &20.8                       &21.0    &20.6    &21.4        &21.8            &29.6\\
oh-\_ch3f\_hoch3fts              &-1.6                       &-1.7    &-0.2    &2.2         &5.1             &-2.7\\
ch3oh\_f-\_hoch3fts              &18.2                       &19.3    &20.9    &22.9        &21.6            &17.6\\
hoch3fcomp2\_hoch3fts            &4.1                        &3.4     &4.0     &4.7         &5.5             &11\\
hoch3fcomp1\_hoch3fts            &43.2                       &45.5    &42.2    &42.2        &44.8            &47.7\\
h\_n2\_hn2ts                     &5.2                        &2.3     &6.8     &6.9         &7.3             &14.6\\
hn2\_hn2ts                       &8.8                        &9.0     &8.7     &9.0         &8.6             &10.9\\
h\_co\_hcots                     &-1.8                       &-3.9    &-0.7    &-0.5        &-0.3            &3.2\\
hco\_hcots                       &24.6                       &25.3    &24.2    &24.3        &24.3            &22.8\\
h\_c2h4\_c2h5ts                  &-0.2                       &-2.0    &0.8     &0.9         &1.5             &2\\
c2h5\_c2h5ts                     &40.2                       &40.3    &40.0    &40.9        &40.0            &42\\
ch3\_c2h4\_c3h7ts                &1.6                        &-1.4    &2.9     &4.7         &4.2             &6.4\\
c3h7\_c3h7ts                     &29.9                       &32.2    &28.8    &28.4        &29.9            &33\\
hcn\_hcnts                       &45.4                       &44.4    &45.9    &45.3        &46.3            &48.1\\
hnc\_hcnts                       &30.5                       &30.1    &30.8    &30.4        &30.9            &33\\
h\_hcl\_RKT01                    &0.2                        &-0.5    &0.8     &0.4         &1.6             &6.1\\
H2\_cl\_RKT01                    &-1.6                       &-4.7    &-0.2    &1.7         &0.3             &8\\
oh\_H2\_RKT02                    &-9.7                       &-13.2   &-8.2    &-6.0        &-7.5            &5.2\\
H2O\_h\_RKT02                    &13.4                       &13.0    &13.7    &13.0        &15.2            &21.6\\
ch3\_H2\_RKT03                   &3.8                        &-0.2    &5.6     &7.4         &5.6             &11.9\\
CH4\_h\_RKT03                    &9.3                        &7.8     &10.2    &9.4         &11.0            &15\\
oh\_CH4\_RKT04                   &-8.8                       &-12.0   &-7.4    &-5.4        &-6.5            &6.3\\
H2O\_ch3\_RKT04                  &8.8                        &6.2     &9.9     &11.6        &10.8            &19.5\\
h\_H2\_RKT06                     &3.6                        &1.1     &5.0     &4.9         &5.1             &9.7\\
h\_H2\_RKT06                     &3.6                        &1.1     &5.0     &4.9         &5.1             &9.7\\
oh\_NH3\_RKT07                   &-15.1                      &-18.5   &-13.7   &-11.4       &-12.7           &3.4\\
H2O\_NH2\_RKT07                  &-0.7                       &-4.6    &1.0     &3.3         &1.8             &13.7\\
hcl\_ch3\_RKT08                  &-5.9                       &-8.9    &-4.5    &-2.8        &-4.1            &1.8\\
cl\_CH4\_RKT08                   &-2.2                       &-5.2    &-0.9    &0.5         &0.0             &6.8\\
oh\_C2H6\_RKT09                  &-12.2                      &-15.5   &-10.7   &-8.6        &-9.8            &3.5\\
H2O\_C2H5\_RKT09                 &10.7                       &8.5     &11.7    &13.3        &12.7            &20.4\\
f\_H2\_RKT10                     &-16.9                      &-18.7   &-16.2   &-14.0       &-15.0           &1.6\\
hf\_h\_RKT10                     &24.1                       &25.4    &23.7    &22.7        &26.0            &33.8\\
O\_CH4\_RKT11                    &-3.0                       &-6.1    &-1.7    &0.6         &-1.0            &14.4\\
oh\_ch3\_RKT11                   &-0.2                       &-3.8    &1.4     &2.7         &1.6             &8.9\\
h\_PH3\_RKT12                    &-1.8                       &-3.8    &-0.7    &-0.9        &-0.5            &2.9\\
H2\_PH2\_RKT12                   &18.4                       &14.5    &20.0    &22.3        &20.3            &24.7\\
h\_oh\_RKT14                     &3.9                        &2.0     &5.0     &4.2         &5.3             &10.9\\
H2\_O\_RKT14                     &-4.4                       &-8.3    &-2.7    &0.1         &-2.7            &13.2\\
h\_H2S\_RKT16                    &-1.3                       &-3.2    &-0.2    &-0.3        &0.0             &3.9\\
H2\_HS\_RKT16                    &8.7                        &4.7     &10.3    &12.5        &10.3            &17.2\\
O\_hcl\_RKT17                    &-16.9                      &-19.4   &-15.9   &-13.3       &-14.5           &10.4\\
oh\_cl\_RKT17                    &-10.4                      &-13.3   &-9.2    &-7.9        &-7.8            &9.9\\
NH2\_ch3\_RKT18                  &0.8                        &-3.0    &2.6     &4.0         &2.8             &8.9\\
NH\_CH4\_RKT18                   &10.7                       &7.4     &12.2    &14.2        &12.9            &22\\
NH2\_C2H5\_RKT19                 &3.1                        &-0.6    &4.8     &6.1         &5.1             &9.8\\
NH\_C2H6\_RKT19                  &7.7                        &4.2     &9.3     &11.4        &10.1            &19.4\\
C2H6\_NH2\_RKT20                 &1.6                        &-3.0    &3.8     &5.8         &4.0             &11.3\\
C2H5\_NH3\_RKT20                 &10.1                       &7.1     &11.5    &13.0        &12.0            &17.8\\
NH2\_CH4\_RKT21                  &4.5                        &0.1     &6.6     &8.5         &6.8             &13.9\\
NH3\_ch3\_RKT21                  &7.8                        &4.6     &9.2     &10.7        &9.7             &16.9\\
C5H8\_RKT22                      &31.1                       &27.3    &32.9    &33.7        &33.1            &39.7\\
C5H8\_RKT22                      &31.1                       &27.3    &32.9    &33.7        &33.1            &39.7\\
ME                               &-8.9                       &-10.7   &-7.7    &-6.6        &-6.9            &~\\
MAE                              &9.1                        &11.2    &8.4     &7.7         &7.8             &~\\\hline
\end{longtable}

\begin{table}[tbp]
    \caption{Self-consistent lattice constants (in \AA) of metal bulks used for constructing the slabs, obtained from seven-point SJEOS fits around the equilibrium volume.}\label{tab:lc-chemisorption}
    \centering
\begin{ruledtabular}
    \begin{tabular}{lcccccc}
        xc & Pt & Pd & Rh & Cu & Ni & Ir \\ \hline
        RPBE & 3.981 & 3.973 & 3.875 & 3.672 & 3.546 & 3.889 \\
        SCAN & 3.888 & 3.908 & 3.805 & 3.559 & 3.482 & 3.796 \\
        $\vartheta$-PBE & 3.954 & 3.939 & 3.857 & 3.646 & 3.510 & 3.873 \\
    \end{tabular}
    \end{ruledtabular}
\end{table}
\endgroup

\begin{figure*}[tbp]
  \centering
  \includegraphics[width=0.96\textwidth]{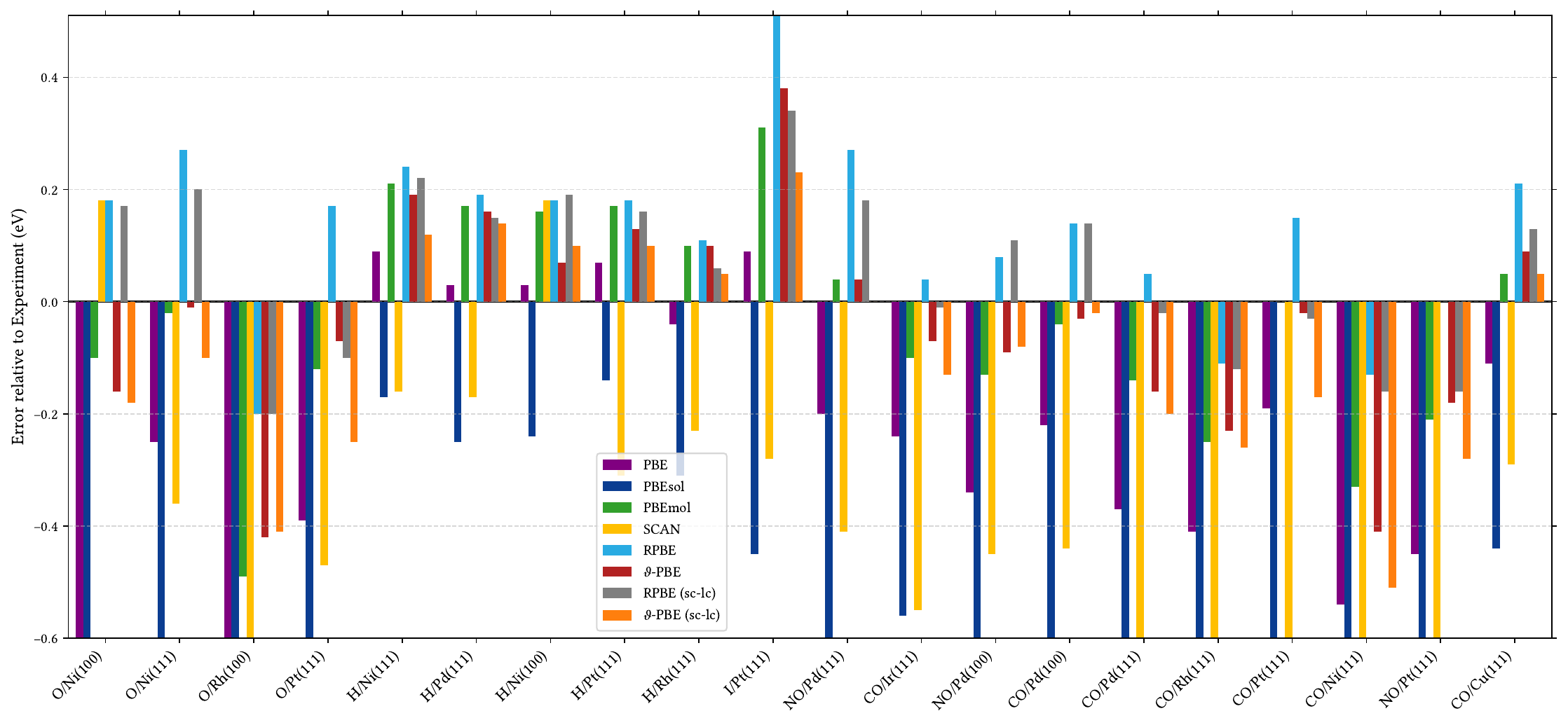}
  \caption{
   Errors in calculated chemisorption energies (in eV) relative to experimental values across selected adsorbate/metal systems. 
   The bar chart compares the performance of $\vartheta$-PBE against various standard density functionals. 
   The data includes both self-consistent lattice evaluations (labeled sc-lc) for RPBE and $\vartheta$-PBE, as well as fixed-lattice benchmark results 
   where the bulk lattice constant is constrained to the SCAN equilibrium value.
  }
  \label{fig:chemerr}
\end{figure*}

\begin{figure*}[ht]
    \centering
        \includegraphics[width=0.4\textwidth]{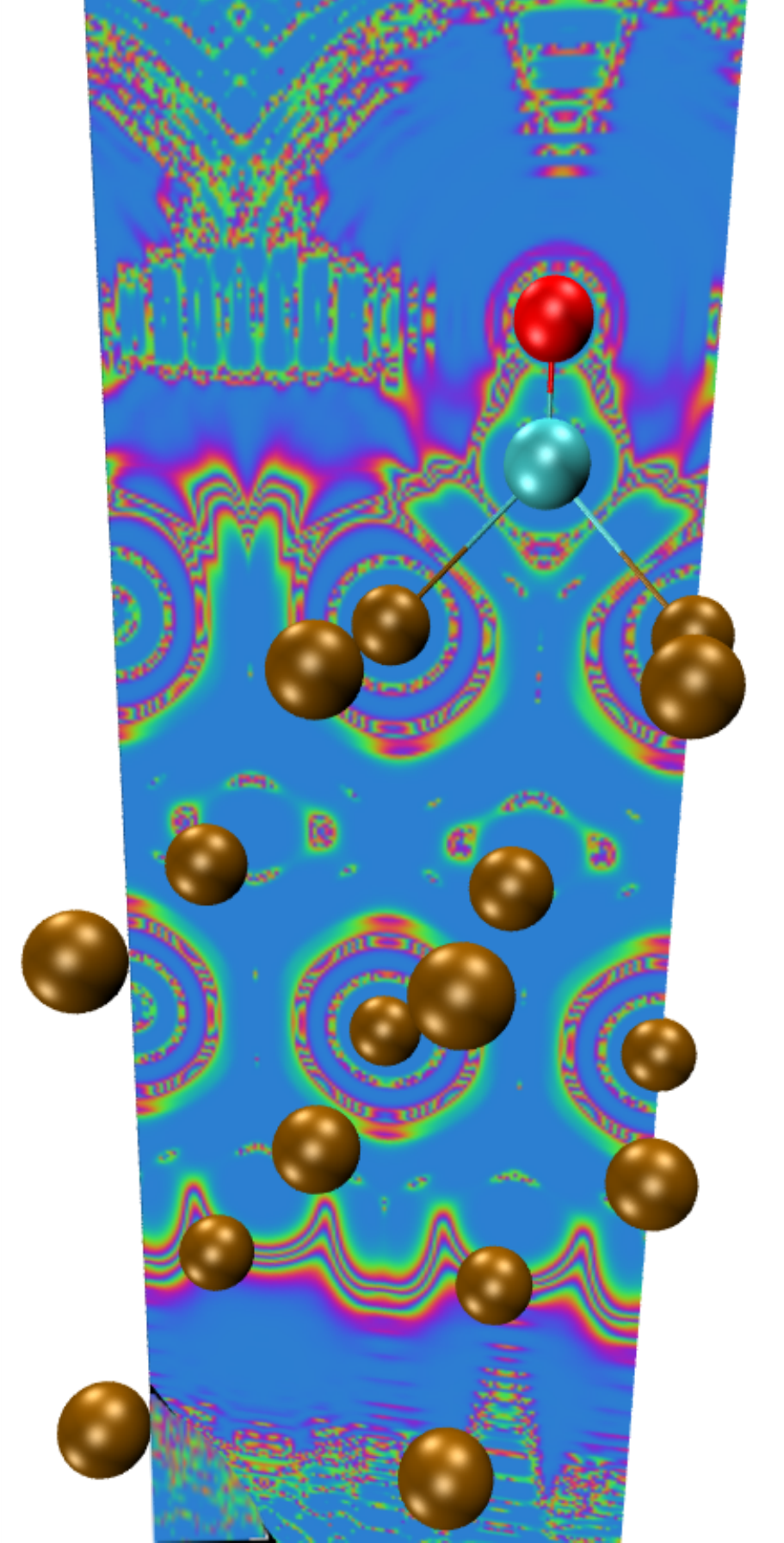}
        \label{fig:copd100}
    \caption{Volume slice visualization of the switching function for a CO/Pd(111).
        The more blue color inidicates $f(\vartheta) \rightarrow 0$ and the more red indicates $f(\vartheta) \rightarrow 1$.
    }
    \label{fig:tvis}

\end{figure*}

%% file: references.bib
@article{ks1965,
  title={Self-consistent equations including exchange and correlation effects},
  author={Kohn, Walter and Sham, Lu Jeu},
  journal={Physical review},
  volume={140},
  number={4A},
  pages={A1133},
  year={1965},
  publisher={APS}
}

@article{hk1964,
  title={Inhomogeneous electron gas},
  author={Hohenberg, Pierre and Kohn, Walter},
  journal={Physical review},
  volume={136},
  number={3B},
  pages={B864},
  year={1964},
  publisher={APS}
}

@article{pbe1996,
  title={Generalized gradient approximation made simple},
  author={Perdew, John P and Burke, Kieron and Ernzerhof, Matthias},
  journal={Physical review letters},
  volume={77},
  number={18},
  pages={3865},
  year={1996},
  publisher={APS}
}

@article{rpbe1999,
  title={Improved adsorption energetics within density-functional theory using revised Perdew-Burke-Ernzerhof functionals},
  author={Hammer, BHLB and Hansen, Lars Bruno and N{\o}rskov, Jens Kehlet},
  journal={Physical review B},
  volume={59},
  number={11},
  pages={7413},
  year={1999},
  publisher={APS}
}

@article{pbeint,
  title={Generalized gradient approximation bridging the rapidly and slowly varying density regimes: A PBE-like functional for hybrid interfaces},
  author={Fabiano, Eduardo and Constantin, Lucian A and Della Sala, Fabio},
  journal={Physical Review B—Condensed Matter and Materials Physics},
  volume={82},
  number={11},
  pages={113104},
  year={2010},
  publisher={APS}
}

@article{pbemol,
  title={Non-empirical improvement of PBE and its hybrid PBE0 for general description of molecular properties},
  author={del Campo, Jorge M and G{\'a}zquez, Jos{\'e} L and Trickey, SB and Vela, Alberto},
  journal={The Journal of chemical physics},
  volume={136},
  number={10},
  year={2012},
  publisher={AIP Publishing}
}

@article{tpss2003,
  title={Climbing the density functional ladder: Nonempirical meta--generalized gradient approximation designed for molecules and solids},
  author={Tao, Jianmin and Perdew, John P and Staroverov, Viktor N and Scuseria, Gustavo E},
  journal={Physical review letters},
  volume={91},
  number={14},
  pages={146401},
  year={2003},
  publisher={APS}
}

@article{scan2015,
  title={Strongly constrained and appropriately normed semilocal density functional},
  author={Sun, Jianwei and Ruzsinszky, Adrienn and Perdew, John P},
  journal={Physical review letters},
  volume={115},
  number={3},
  pages={036402},
  year={2015},
  publisher={APS}
}

@article{r2scan2020,
  title={Accurate and numerically efficient r2SCAN meta-generalized gradient approximation},
  author={Furness, James W and Kaplan, Aaron D and Ning, Jinliang and Perdew, John P and Sun, Jianwei},
  journal={The journal of physical chemistry letters},
  volume={11},
  number={19},
  pages={8208--8215},
  year={2020},
  publisher={ACS Publications}
}

@Article{dori,
author={de Silva, Piotr
and Corminboeuf, Cl{\'e}mence},
title={Simultaneous Visualization of Covalent and Noncovalent Interactions Using Regions of Density Overlap},
journal={Journal of Chemical Theory and Computation},
year={2014},
month={Sep},
day={09},
publisher={American Chemical Society},
volume={10},
number={9},
pages={3745-3756},
issn={1549-9618},
doi={10.1021/ct500490b},
url={https://doi.org/10.1021/ct500490b}
}

@article{thetamgga,
  title={Communication: A new class of non-empirical explicit density functionals on the third rung of Jacob’s ladder},
  author={{de Silva}, Piotr and Corminboeuf, Cl{\'e}mence},
  journal={The Journal of Chemical Physics},
  volume={143},
  number={11},
  pages={111105},
  year={2015},
  publisher={AIP Publishing LLC}
}

@article{sedd,
    author = {de Silva, Piotr and Korchowiec, Jacek and Wesolowski, Tomasz A.},
    title = {Atomic shell structure from the Single-Exponential Decay Detector},
    journal = {The Journal of Chemical Physics},
    volume = {140},
    number = {16},
    pages = {164301},
    year = {2014},
    month = {04},
    issn = {0021-9606},
    doi = {10.1063/1.4871501},
    url = {https://doi.org/10.1063/1.4871501},
    eprint = {https://pubs.aip.org/aip/jcp/article-pdf/doi/10.1063/1.4871501/14006898/164301\_1\_online.pdf},
}

@article{desilva2012revealing,
  title={Revealing the bonding pattern from the molecular electron density using single exponential decay detector: an orbital-free alternative to the electron localization function},
  author={de Silva, Piotr and Korchowiec, Jacek and Wesolowski, Tomasz A},
  journal={ChemPhysChem},
  volume={13},
  number={15},
  pages={3462--3465},
  year={2012},
  publisher={Wiley Online Library}
}

@article{desilva2013extracting,
  title={Extracting information about chemical bonding from molecular electron densities via single exponential decay detector (SEDD)},
  author={de Silva, Piotr and Korchowiec, Jacek and JS, Nirmal Ram and Wesolowski, Tomasz A},
  journal={Chimia},
  volume={67},
  number={4},
  pages={253--253},
  year={2013}
}

@Article{pig,
author ="Maier, Toni M. and Haasler, Matthias and Arbuznikov, Alexei V. and Kaupp, Martin",
title  ="New approaches for the calibration of exchange-energy densities in local hybrid functionals",
journal  ="Phys. Chem. Chem. Phys.",
year  ="2016",
volume  ="18",
issue  ="31",
pages  ="21133-21144",
publisher  ="The Royal Society of Chemistry",
doi  ="10.1039/C6CP00990E",
url  ="http://dx.doi.org/10.1039/C6CP00990E"}

@article{mcbook,
  title={The matrix cookbook},
  author={Petersen, Kaare Brandt and Pedersen, Michael Syskind and others},
  journal={Technical University of Denmark},
  volume={7},
  number={15},
  pages={510},
  year={2008}
}

@article{blochl-paw,
  title = {Projector augmented-wave method},
  author = {Bl\"ochl, P. E.},
  journal = {Phys. Rev. B},
  volume = {50},
  issue = {24},
  pages = {17953--17979},
  numpages = {0},
  year = {1994},
  month = {Dec},
  publisher = {American Physical Society},
  doi = {10.1103/PhysRevB.50.17953},
  url = {https://link.aps.org/doi/10.1103/PhysRevB.50.17953}
}

@article{blochl2003paw,
  title={Projector augmented wave method: ab initio molecular dynamics with full wave functions},
  author={Bl{\"o}chl, Peter E and F{\"o}rst, Clemens J and Schimpl, Johannes},
  journal={Bulletin of Materials Science},
  volume={26},
  number={1},
  pages={33--41},
  year={2003},
  publisher={Springer}
}

@article{pawvasp,
  title = {From ultrasoft pseudopotentials to the projector augmented-wave method},
  author = {Kresse, G. and Joubert, D.},
  journal = {Phys. Rev. B},
  volume = {59},
  issue = {3},
  pages = {1758--1775},
  numpages = {0},
  year = {1999},
  month = {Jan},
  publisher = {American Physical Society},
  doi = {10.1103/PhysRevB.59.1758},
  url = {https://link.aps.org/doi/10.1103/PhysRevB.59.1758}
}

@article{fd-1994,
  title = {Higher-order finite-difference pseudopotential method: An application to diatomic molecules},
  author = {Chelikowsky, James R. and Troullier, N. and Wu, K. and Saad, Y.},
  journal = {Phys. Rev. B},
  volume = {50},
  issue = {16},
  pages = {11355--11364},
  numpages = {0},
  year = {1994},
  month = {Oct},
  publisher = {American Physical Society},
  doi = {10.1103/PhysRevB.50.11355},
  url = {https://link.aps.org/doi/10.1103/PhysRevB.50.11355}
}

@misc{paw2010,
  doi = {10.48550/ARXIV.0910.1921},
  url = {https://arxiv.org/abs/0910.1921},
  author = {Rostgaard,  Carsten},
  title = {The Projector Augmented-wave Method},
  publisher = {arXiv},
  year = {2009},
  copyright = {arXiv.org perpetual,  non-exclusive license}
}

@article{thetahyb2015,
  title={Local hybrid functionals with orbital-free mixing functions and balanced elimination of self-interaction error},
  author={De Silva, Piotr and Corminboeuf, Cl{\'e}mence},
  journal={The Journal of chemical physics},
  volume={142},
  number={7},
  year={2015},
  publisher={AIP Publishing}
}

@article{perdew85,
  title = {Accurate Density Functional for the Energy: Real-Space Cutoff of the Gradient Expansion for the Exchange Hole},
  author = {Perdew, John P.},
  journal = {Phys. Rev. Lett.},
  volume = {55},
  issue = {16},
  pages = {1665--1668},
  numpages = {0},
  year = {1985},
  month = {Oct},
  publisher = {American Physical Society},
  doi = {10.1103/PhysRevLett.55.1665},
  url = {https://link.aps.org/doi/10.1103/PhysRevLett.55.1665}
}

@article{pw86,
  title = {Accurate and simple density functional for the electronic exchange energy: Generalized gradient approximation},
  author = {Perdew, John P. and Yue, Wang},
  journal = {Phys. Rev. B},
  volume = {33},
  issue = {12},
  pages = {8800--8802},
  numpages = {0},
  year = {1986},
  month = {Jun},
  publisher = {American Physical Society},
  doi = {10.1103/PhysRevB.33.8800},
  url = {https://link.aps.org/doi/10.1103/PhysRevB.33.8800}
}

@article{lda_pw1992,
  title = {Accurate and simple analytic representation of the electron-gas correlation energy},
  author = {Perdew, John P. and Wang, Yue},
  journal = {Phys. Rev. B},
  volume = {45},
  issue = {23},
  pages = {13244--13249},
  numpages = {0},
  year = {1992},
  month = {Jun},
  publisher = {American Physical Society},
  doi = {10.1103/PhysRevB.45.13244},
  url = {https://link.aps.org/doi/10.1103/PhysRevB.45.13244}
}

@article{lsda_vonbarth1972,
doi = {10.1088/0022-3719/5/13/012},
url = {https://dx.doi.org/10.1088/0022-3719/5/13/012},
year = {1972},
month = {Jul},
publisher = {},
volume = {5},
number = {13},
pages = {1629},
author = {U von Barth and L Hedin},
title = {A local exchange-correlation potential for the spin polarized case. i},
journal = {Journal of Physics C: Solid State Physics}
}

@article{mgga_ghoshparr86,
  title = {Phase-space approach to the exchange-energy functional of density-functional theory},
  author = {Ghosh, Swapan K. and Parr, Robert G.},
  journal = {Phys. Rev. A},
  volume = {34},
  issue = {2},
  pages = {785--791},
  numpages = {0},
  year = {1986},
  month = {Aug},
  publisher = {American Physical Society},
  doi = {10.1103/PhysRevA.34.785},
  url = {https://link.aps.org/doi/10.1103/PhysRevA.34.785}
}

@article{pbesol,
  title = {Restoring the Density-Gradient Expansion for Exchange in Solids and Surfaces},
  author = {Perdew, John P. and Ruzsinszky, Adrienn and Csonka, G\'abor I. and Vydrov, Oleg A. and Scuseria, Gustavo E. and Constantin, Lucian A. and Zhou, Xiaolan and Burke, Kieron},
  journal = {Phys. Rev. Lett.},
  volume = {100},
  issue = {13},
  pages = {136406},
  numpages = {4},
  year = {2008},
  month = {Apr},
  publisher = {American Physical Society},
  doi = {10.1103/PhysRevLett.100.136406},
  url = {https://link.aps.org/doi/10.1103/PhysRevLett.100.136406}
}

@article{vs1998,
    author = {Van Voorhis, Troy and Scuseria, Gustavo E.},
    title = {A novel form for the exchange-correlation energy functional},
    journal = {The Journal of Chemical Physics},
    volume = {109},
    number = {2},
    pages = {400-410},
    year = {1998},
    month = {07},
    issn = {0021-9606},
    doi = {10.1063/1.476577},
    url = {https://doi.org/10.1063/1.476577},
    eprint = {https://pubs.aip.org/aip/jcp/article-pdf/109/2/400/19044609/400\_1\_online.pdf},
}

@article{gpaw2024,
    author = {Mortensen, Jens Jørgen and Larsen, Ask Hjorth and Kuisma, Mikael and Ivanov, Aleksei V. and Taghizadeh, Alireza and Peterson, Andrew and Haldar, Anubhab and Dohn, Asmus Ougaard and Schäfer, Christian and Jónsson, Elvar Örn and Hermes, Eric D. and Nilsson, Fredrik Andreas and Kastlunger, Georg and Levi, Gianluca and Jónsson, Hannes and Häkkinen, Hannu and Fojt, Jakub and Kangsabanik, Jiban and Sødequist, Joachim and Lehtomäki, Jouko and Heske, Julian and Enkovaara, Jussi and Winther, Kirsten Trøstrup and Dulak, Marcin and Melander, Marko M. and Ovesen, Martin and Louhivuori, Martti and Walter, Michael and Gjerding, Morten and Lopez-Acevedo, Olga and Erhart, Paul and Warmbier, Robert and Würdemann, Rolf and Kaappa, Sami and Latini, Simone and Boland, Tara Maria and Bligaard, Thomas and Skovhus, Thorbjørn and Susi, Toma and Maxson, Tristan and Rossi, Tuomas and Chen, Xi and Schmerwitz, Yorick Leonard A. and Schiøtz, Jakob and Olsen, Thomas and Jacobsen, Karsten Wedel and Thygesen, Kristian Sommer},
    title = {GPAW: An open Python package for electronic structure calculations},
    journal = {The Journal of Chemical Physics},
    volume = {160},
    number = {9},
    pages = {092503},
    year = {2024},
    month = {03},
    issn = {0021-9606},
    doi = {10.1063/5.0182685},
    url = {https://doi.org/10.1063/5.0182685},
    eprint = {https://pubs.aip.org/aip/jcp/article-pdf/doi/10.1063/5.0182685/19717263/092503\_1\_5.0182685.pdf},
}

@article{neumann1997higher,
  title={Higher-order gradient corrections for exchange-correlation functionals},
  author={Neumann, Ralf and Handy, Nicholas C},
  journal={Chemical physics letters},
  volume={266},
  number={1-2},
  pages={16--22},
  year={1997},
  publisher={Elsevier}
}

@article{revtpss,
  title = {Workhorse Semilocal Density Functional for Condensed Matter Physics and Quantum Chemistry},
  author = {Perdew, John P. and Ruzsinszky, Adrienn and Csonka, G\'abor I. and Constantin, Lucian A. and Sun, Jianwei},
  journal = {Phys. Rev. Lett.},
  volume = {103},
  issue = {2},
  pages = {026403},
  numpages = {4},
  year = {2009},
  month = {Jul},
  publisher = {American Physical Society},
  doi = {10.1103/PhysRevLett.103.026403},
  url = {https://link.aps.org/doi/10.1103/PhysRevLett.103.026403}
}

@article{spinscaling-perdew,
  title = {Spin-density gradient expansion for the kinetic energy},
  author = {Oliver, G. L. and Perdew, J. P.},
  journal = {Phys. Rev. A},
  volume = {20},
  issue = {2},
  pages = {397--403},
  numpages = {0},
  year = {1979},
  month = {Aug},
  publisher = {American Physical Society},
  doi = {10.1103/PhysRevA.20.397},
  url = {https://link.aps.org/doi/10.1103/PhysRevA.20.397}
}

@article{h2plus-sie,
  title = {Inapplicability of exact constraints and a minimal two-parameter generalization to the DFT+U based correction of self-interaction error},
  author = {Moynihan, Glenn and Teobaldi, Gilberto and O'Regan, David D.},
  journal = {Phys. Rev. B},
  volume = {94},
  issue = {22},
  pages = {220104},
  numpages = {6},
  year = {2016},
  month = {Dec},
  publisher = {American Physical Society},
  doi = {10.1103/PhysRevB.94.220104},
  url = {https://link.aps.org/doi/10.1103/PhysRevB.94.220104}
}

@article{adsorption_validation,
author = {Duanmu, Kaining and Truhlar, Donald G.},
title = {Validation of Density Functionals for Adsorption Energies on Transition Metal Surfaces},
journal = {Journal of Chemical Theory and Computation},
volume = {13},
number = {2},
pages = {835-842},
year = {2017},
doi = {10.1021/acs.jctc.6b01156},
note ={PMID: 27983852},
URL = { https://doi.org/10.1021/acs.jctc.6b01156 },
eprint = { https://doi.org/10.1021/acs.jctc.6b01156 }
}

@article{rtpss2018,
author = {Garza, Alejandro
J. and Bell, Alexis T. and Head-Gordon, Martin},
title = {Nonempirical Meta-Generalized Gradient Approximations for Modeling Chemisorption at Metal Surfaces},
journal = {Journal of Chemical Theory and Computation},
volume = {14},
number = {6},
pages = {3083-3090},
year = {2018},
doi = {10.1021/acs.jctc.8b00288},
note ={PMID: 29746113},
URL = { https://doi.org/10.1021/acs.jctc.8b00288 },
eprint = { https://doi.org/10.1021/acs.jctc.8b00288 }
}

@article{ad_wellendorff2015,
title = {A benchmark database for adsorption bond energies to transition metal surfaces and comparison to selected DFT functionals},
journal = {Surface Science},
volume = {640},
pages = {36-44},
year = {2015},
note = {Reactivity Concepts at Surfaces: Coupling Theory with Experiment},
issn = {0039-6028},
doi = {https://doi.org/10.1016/j.susc.2015.03.023},
url = {https://www.sciencedirect.com/science/article/pii/S0039602815000837},
author = {Jess Wellendorff and Trent L. Silbaugh and Delfina Garcia-Pintos and Jens K. Nørskov and Thomas Bligaard and Felix Studt and Charles T. Campbell}
}

@article{araujo2022adsorption,
  title={Adsorption energies on transition metal surfaces: towards an accurate and balanced description},
  author={Araujo, Rafael B and Rodrigues, Gabriel LS and Dos Santos, Egon Campos and Pettersson, Lars GM},
  journal={Nature Communications},
  volume={13},
  number={1},
  pages={6853},
  year={2022},
  publisher={Nature Publishing Group UK London}
}

@article{becke88,
  title = {Density-functional exchange-energy approximation with correct asymptotic behavior},
  author = {Becke, A. D.},
  journal = {Phys. Rev. A},
  volume = {38},
  issue = {6},
  pages = {3098--3100},
  numpages = {0},
  year = {1988},
  month = {Sep},
  publisher = {American Physical Society},
  doi = {10.1103/PhysRevA.38.3098},
  url = {https://link.aps.org/doi/10.1103/PhysRevA.38.3098}
}

@article{pbeparams,
  title = {Systematic investigation of a family of gradient-dependent functionals for solids},
  author = {Haas, Philipp and Tran, Fabien and Blaha, Peter and Pedroza, Luana S. and da Silva, Antonio J. R. and Odashima, Mariana M. and Capelle, Klaus},
  journal = {Phys. Rev. B},
  volume = {81},
  issue = {12},
  pages = {125136},
  numpages = {10},
  year = {2010},
  month = {Mar},
  publisher = {American Physical Society},
  doi = {10.1103/PhysRevB.81.125136},
  url = {https://link.aps.org/doi/10.1103/PhysRevB.81.125136}
}

@article{becke1989_mgga,
  title = {Exchange holes in inhomogeneous systems: A coordinate-space model},
  author = {Becke, A. D. and Roussel, M. R.},
  journal = {Phys. Rev. A},
  volume = {39},
  issue = {8},
  pages = {3761--3767},
  numpages = {0},
  year = {1989},
  month = {Apr},
  publisher = {American Physical Society},
  doi = {10.1103/PhysRevA.39.3761},
  url = {https://link.aps.org/doi/10.1103/PhysRevA.39.3761}
}

@article{LAK2024,
  title = {Balancing the Contributions to the Gradient Expansion: Accurate Binding and Band Gaps with a Nonempirical Meta-GGA},
  author = {Lebeda, Timo and Aschebrock, Thilo and K\"ummel, Stephan},
  journal = {Phys. Rev. Lett.},
  volume = {133},
  issue = {13},
  pages = {136402},
  numpages = {9},
  year = {2024},
  month = {Sep},
  publisher = {American Physical Society},
  doi = {10.1103/PhysRevLett.133.136402},
  url = {https://link.aps.org/doi/10.1103/PhysRevLett.133.136402}
}

@article{scanl,
  title = {Deorbitalization strategies for meta-generalized-gradient-approximation exchange-correlation functionals},
  author = {Mejia-Rodriguez, Daniel and Trickey, S. B.},
  journal = {Phys. Rev. A},
  volume = {96},
  issue = {5},
  pages = {052512},
  numpages = {10},
  year = {2017},
  month = {Nov},
  publisher = {American Physical Society},
  doi = {10.1103/PhysRevA.96.052512},
  url = {https://link.aps.org/doi/10.1103/PhysRevA.96.052512}
}

@article{ofr2,
  title = {Laplacian-level meta-generalized gradient approximation for solid and liquid metals},
  author = {Kaplan, Aaron D. and Perdew, John P.},
  journal = {Phys. Rev. Mater.},
  volume = {6},
  issue = {8},
  pages = {083803},
  numpages = {32},
  year = {2022},
  month = {Aug},
  publisher = {American Physical Society},
  doi = {10.1103/PhysRevMaterials.6.083803},
  url = {https://link.aps.org/doi/10.1103/PhysRevMaterials.6.083803}
}

@article{scanl-bench,
  title = {Deorbitalized meta-GGA exchange-correlation functionals in solids},
  author = {Mejia-Rodriguez, Daniel and Trickey, S. B.},
  journal = {Phys. Rev. B},
  volume = {98},
  issue = {11},
  pages = {115161},
  numpages = {9},
  year = {2018},
  month = {Sep},
  publisher = {American Physical Society},
  doi = {10.1103/PhysRevB.98.115161},
  url = {https://link.aps.org/doi/10.1103/PhysRevB.98.115161}
}

@article{von1935,
  title={Phys. 1935 96 431 Von Weiz{\"a}cker, CF},
  author={Von Weiz{\"a}cker, CFZ},
  journal={Z. Phys},
  volume={96},
  pages={431},
  year={1935}
}

@article{fermi1927,
  title={Application of statistical gas methods to electronic systems},
  author={Fermi, E},
  journal={Atti accad. naz. Lincei},
  volume={6},
  pages={602--607},
  year={1927}
}

@article{r2scanl,
  title = {Meta-GGA performance in solids at almost GGA cost},
  author = {Mej\'{\i}a-Rodr\'{\i}guez, Daniel and Trickey, S. B.},
  journal = {Phys. Rev. B},
  volume = {102},
  issue = {12},
  pages = {121109},
  numpages = {4},
  year = {2020},
  month = {Sep},
  publisher = {American Physical Society},
  doi = {10.1103/PhysRevB.102.121109},
  url = {https://link.aps.org/doi/10.1103/PhysRevB.102.121109}
}

@article{ae6bh6,
  title={Small representative benchmarks for thermochemical calculations},
  author={Lynch, Benjamin J and Truhlar, Donald G},
  journal={The Journal of Physical Chemistry A},
  volume={107},
  number={42},
  pages={8996--8999},
  year={2003},
  publisher={ACS Publications}
}

@article{ae6bh6ref,
  title={Theoretical reference values for the AE6 and BH6 test sets from explicitly correlated coupled-cluster theory},
  author={Haunschild, Robin and Klopper, Wim},
  journal={Theoretical Chemistry Accounts},
  volume={131},
  number={2},
  pages={1112},
  year={2012},
  publisher={Springer}
}

@article{scfmgga_2011,
  title = {Self-consistent meta-generalized gradient approximation within the projector-augmented-wave method},
  author = {Sun, Jianwei and Marsman, Martijn and Csonka, G\'abor I. and Ruzsinszky, Adrienn and Hao, Pan and Kim, Yoon-Suk and Kresse, Georg and Perdew, John P.},
  journal = {Phys. Rev. B},
  volume = {84},
  issue = {3},
  pages = {035117},
  numpages = {12},
  year = {2011},
  month = {Jul},
  publisher = {American Physical Society},
  doi = {10.1103/PhysRevB.84.035117},
  url = {https://link.aps.org/doi/10.1103/PhysRevB.84.035117}
}

@article{lieb,
  title={Improved lower bound on the indirect Coulomb energy},
  author={Lieb, Elliott H and Oxford, Stephen},
  journal={International Journal of Quantum Chemistry},
  volume={19},
  number={3},
  pages={427--439},
  year={1981},
  publisher={Wiley Online Library}
}

@misc{perdew1991electronic,
  title={Electronic structure of solids’ 91},
  author={Perdew, John P and Ziesche, P and Eschrig, H},
  year={1991},
  publisher={Akademie Verlag Berlin}
}

@article{gks,
  title = {Generalized Kohn-Sham schemes and the band-gap problem},
  author = {Seidl, A. and G\"orling, A. and Vogl, P. and Majewski, J. A. and Levy, M.},
  journal = {Phys. Rev. B},
  volume = {53},
  issue = {7},
  pages = {3764--3774},
  numpages = {0},
  year = {1996},
  month = {Feb},
  publisher = {American Physical Society},
  doi = {10.1103/PhysRevB.53.3764},
  url = {https://link.aps.org/doi/10.1103/PhysRevB.53.3764}
}

@article{sjeos,
  title = {Energy and pressure versus volume: Equations of state motivated by the stabilized jellium model},
  author = {Alchagirov, Alim B. and Perdew, John P. and Boettger, Jonathan C. and Albers, R. C. and Fiolhais, Carlos},
  journal = {Phys. Rev. B},
  volume = {63},
  issue = {22},
  pages = {224115},
  numpages = {16},
  year = {2001},
  month = {May},
  publisher = {American Physical Society},
  doi = {10.1103/PhysRevB.63.224115},
  url = {https://link.aps.org/doi/10.1103/PhysRevB.63.224115}
}

@article{ase-paper,
  author={Ask Hjorth Larsen and Jens Jørgen Mortensen and Jakob Blomqvist and Ivano E Castelli and Rune Christensen and Marcin
Dułak and Jesper Friis and Michael N Groves and Bjørk Hammer and Cory Hargus and Eric D Hermes and Paul C Jennings and Peter
Bjerre Jensen and James Kermode and John R Kitchin and Esben Leonhard Kolsbjerg and Joseph Kubal and Kristen
Kaasbjerg and Steen Lysgaard and Jón Bergmann Maronsson and Tristan Maxson and Thomas Olsen and Lars Pastewka and Andrew
Peterson and Carsten Rostgaard and Jakob Schiøtz and Ole Schütt and Mikkel Strange and Kristian S Thygesen and Tejs
Vegge and Lasse Vilhelmsen and Michael Walter and Zhenhua Zeng and Karsten W Jacobsen},
  title={The atomic simulation environment—a Python library for working with atoms},
  journal={Journal of Physics: Condensed Matter},
  volume={29},
  number={27},
  pages={273002},
  url={http://stacks.iop.org/0953-8984/29/i=27/a=273002},
  year={2017}
}

@article{monkhorst-pack,
  title = {Special points for Brillouin-zone integrations},
  author = {Monkhorst, Hendrik J. and Pack, James D.},
  journal = {Phys. Rev. B},
  volume = {13},
  issue = {12},
  pages = {5188--5192},
  numpages = {0},
  year = {1976},
  month = {Jun},
  publisher = {American Physical Society},
  doi = {10.1103/PhysRevB.13.5188},
  url = {https://link.aps.org/doi/10.1103/PhysRevB.13.5188}
}

@article{methfessel-paxton,
  title = {High-precision sampling for Brillouin-zone integration in metals},
  author = {Methfessel, M. and Paxton, A. T.},
  journal = {Phys. Rev. B},
  volume = {40},
  issue = {6},
  pages = {3616--3621},
  numpages = {0},
  year = {1989},
  month = {Aug},
  publisher = {American Physical Society},
  doi = {10.1103/PhysRevB.40.3616},
  url = {https://link.aps.org/doi/10.1103/PhysRevB.40.3616}
}

@article{mabruckner,
  title={Correlation energy of an electron gas with a slowly varying high density},
  author={BRUECKNER, KEITH A and others},
  journal={Physical Review},
  volume={165},
  number={1},
  pages={18},
  year={1968},
  publisher={APS}
}

@article{gmtkn55,
  title={A look at the density functional theory zoo with the advanced GMTKN55 database for general main group thermochemistry, kinetics and noncovalent interactions},
  author={Goerigk, Lars and Hansen, Andreas and Bauer, Christoph and Ehrlich, Stephan and Najibi, Asim and Grimme, Stefan},
  journal={Physical Chemistry Chemical Physics},
  volume={19},
  number={48},
  pages={32184--32215},
  year={2017},
  publisher={Royal Society of Chemistry}
}

@article{bh76,
  title={Benchmark database of barrier heights for heavy atom transfer, nucleophilic substitution, association, and unimolecular reactions and its use to test theoretical methods},
  author={Zhao, Yan and Gonz{\'a}lez-Garc{\'\i}a, N{\'u}ria and Truhlar, Donald G},
  journal={The Journal of Physical Chemistry A},
  volume={109},
  number={9},
  pages={2012--2018},
  year={2005},
  publisher={ACS Publications}
}

@article{cohen2012challenges,
  title={Challenges for density functional theory},
  author={Cohen, Aron J and Mori-S{\'a}nchez, Paula and Yang, Weitao},
  journal={Chemical reviews},
  volume={112},
  number={1},
  pages={289--320},
  year={2012},
  publisher={ACS Publications}
}

@article{shapeoperator,
  title={The role of shape operator in gauge theories},
  author={Zatloukal, V{\'a}clav and Vedl, {\v{S}}imon},
  journal={International Journal of Geometric Methods in Modern Physics},
  volume={22},
  number={02},
  pages={2450271},
  year={2025},
  publisher={World Scientific}
}

@article{beef-vdw,
  title = {Density functionals for surface science: Exchange-correlation model development with Bayesian error estimation},
  author = {Wellendorff, Jess and Lundgaard, Keld T. and M\o{}gelh\o{}j, Andreas and Petzold, Vivien and Landis, David D. and N\o{}rskov, Jens K. and Bligaard, Thomas and Jacobsen, Karsten W.},
  journal = {Phys. Rev. B},
  volume = {85},
  issue = {23},
  pages = {235149},
  numpages = {23},
  year = {2012},
  month = {Jun},
  publisher = {American Physical Society},
  doi = {10.1103/PhysRevB.85.235149},
  url = {https://link.aps.org/doi/10.1103/PhysRevB.85.235149}
}

@article{dftd,
    author = {Grimme, Stefan and Antony, Jens and Ehrlich, Stephan and Krieg, Helge},
    title = {A consistent and accurate ab initio parametrization of density functional dispersion correction (DFT-D) for the 94 elements H-Pu},
    journal = {The Journal of Chemical Physics},
    volume = {132},
    number = {15},
    pages = {154104},
    year = {2010},
    month = {04},
    issn = {0021-9606},
    doi = {10.1063/1.3382344},
    url = {https://doi.org/10.1063/1.3382344},
    eprint = {https://pubs.aip.org/aip/jcp/article-pdf/doi/10.1063/1.3382344/15684000/154104_1_online.pdf},
}

@article{ge2,
  title = {Gradient expansion of the exchange energy from second-order density response theory},
  author = {Svendsen, P. S. and von Barth, U.},
  journal = {Phys. Rev. B},
  volume = {54},
  issue = {24},
  pages = {17402--17413},
  numpages = {0},
  year = {1996},
  month = {Dec},
  publisher = {American Physical Society},
  doi = {10.1103/PhysRevB.54.17402},
  url = {https://link.aps.org/doi/10.1103/PhysRevB.54.17402}
}

@misc{thetapbe,
  author = {},
  title = {Self-consistent implementation of Hessian-level meta-GGAs},
  date = {2026},
  publisher    = {GitLab},
  journal      = {GitLab Merge Request},
  howpublished = {\url{https://gitlab.com/gpaw/gpaw/-/merge_requests/3232}},
}
